 \documentclass{OldTemplate/jfm}

\usepackage{newtxtext}
\usepackage{newtxmath}
\usepackage{natbib}
\usepackage{graphicx}
\usepackage{epstopdf, epsfig}
\usepackage{dcolumn}
\usepackage{bm}
\usepackage{hyperref}
\hypersetup{pdfborder={0 0 0},
            colorlinks=true,
            linkcolor=blue,
            citecolor=blue,
            urlcolor=blue}

\usepackage{upgreek}
\usepackage{color}
\usepackage{appendix}
\usepackage[section]{placeins}
\usepackage{amsmath,mathtools}
\usepackage{fancyhdr}
\usepackage{acronym}
    \acrodef{PIV}[PIV]{particle image velocimetry}
    \acrodef{LES}[LES]{large eddy simulation}
    \acrodef{DNS}[DNS]{direct numerical simulation}
    \acrodef{POD}[POD]{proper orthogonal decomposition}
    \acrodef{DMD}[DMD]{dynamic mode decomposition}
    \acrodef{SPOD}[SPOD]{spectral proper orthogonal decomposition}
    \acrodef{HPOD}[HPOD]{Hilbert proper orthogonal decomposition}
    \acrodef{DFT}[DFT]{discrete Fourier transform}
     \acrodef{mPOD}[mPOD]{multi-scale proper orthogonal decomposition}
     \acrodef{FFT}[FFT]{fast Fourier transform}
     \acrodef{PSD}[PSD]{power spectral density}
     \acrodef{COEF}[COEF]{complex empirical orthogonal functions}
     \acrodef{CPC}[CPC]{complex principal components}
     \acrodef{COD}[COD]{complex orthogonal decomposition}

\newcommand{\cmt}[1]{\ignorespaces}
\newcommand{\dx}{\mathrm{d}}
\newcommand{\eref}[1]{(\ref{#1})}

 \title{Hilbert Proper Orthogonal Decomposition: a tool for educing advective wavepackets from flow field data}

 \author{Marco Raiola\aff{1,2}
   \corresp{\email{mraiola@ing.uc3m.es}} 
   \and Jochen Kriegseis\aff{2}}
 \affiliation{\aff{1}Aerospace Engineering Department, Universidad Carlos III de Madrid, Legan\'es 28912, Spain
 \aff{2}Institute of Fluid Mechanics, Karlsruhe Institute of Technology, Karlsruhe D-76131, Germany}

\begin{document}

\maketitle

\begin{abstract}
Travelling wavepackets are key coherent features contributing to the dynamics of several advective flows. This work introduces the Hilbert proper orthogonal decomposition (HPOD) to distil these features from flow field data, leveraging their mathematical representation as modulated travelling waves. The HPOD is a complex-valued extension of the proper orthogonal decomposition, where the Hilbert transform of the dataset is used to compute its analytic signal. Two versions of the technique are explored and compared: the conventional HPOD, computing the analytic signal in time; a novel space-only HPOD, computing it along the advection direction. The HPOD is shown to extract wavepackets with amplitude and frequency modulation in time and space. Its broadband nature offers an alternative to spectrally-pure decompositions when instantaneous, local wave characteristics are important. 
The space-only version, leveraging space/time equivalence in travelling waves to swap temporal operations by spatial ones, is proved mathematically equivalent to its conventional counterpart.
The two HPOD versions are characterized and validated on three datasets ordered by complexity: a 2D-DNS of a laminar bluff-body wake with periodic vortex shedding; an LES of a turbulent jet with intermittent, highly modulated wavepackets; and a 2D-PIV of a turbulent jet with measurement errors and no temporal resolution. In advecting flows, both HPOD versions deliver practically identical complex-valued advecting wavepacket structures, characterized by spatiotemporal amplification and decay, wave modulation and intermittency phenomena in turbulent flow cases, such as in turbulent jets. The space-only variant allows to extract these structures from temporally under-resolved datasets, typical of snapshot particle image velocimetry.
\end{abstract}


\section{Introduction}
One of the primary challenges faced in modern fluid dynamics became simplifying and distilling the complex dynamics of fluid flows into a more understandable form; even in seemingly simple configurations, fluid flows often exhibit intricate and unpredictable behaviours. This longstanding challenge stems from the inherent non-linearity and chaotic nature of fluid dynamics, particularly in turbulent flows, where flow structures of various sizes and timescales dynamically interact. A key instrument to reduce the complexity of these high-dimensional flows are modal decomposition techniques, capable of dividing the problem into more manageable low-dimensional flow structures. Comprehensive reviews of these techniques can be found in the works of \cite{rowley2017model} and \cite{taira2017modal}. 

In this work, we propose a data-driven modal decomposition technique specifically designed
to distil spatiotemporal coherent features in advection-dominated flows. These coherent features are, typically, representative of the advective flow structures populating these flows.
The vortex shedding pattern produced by the Bérnand-Von Kármán instability is a clear example of advective flow structure emerging in both laminar or turbulent bluff-body wake flows \citep{perry1982vortex}.
Advecting structures are observed also in shear layers, where vortices are produced through the Kelvin-Helmholtz instability. This case is especially relevant in jet flows, where distinct advecting patterns, typically referred to as wavepackets, have been observed to be a key component of subsonic jet noise \citep{jordan2013wavepackets}. Wall-bounded flows are characterized by advecting structures at different scales: coherent waves are found in the viscous layer near the wall or as very large structures spanning the entire boundary layer, both advecting with uniform velocity ; transient bursts advecting with non-uniform velocity are typically associated with the logarithmic layer \citep{jimenez2018coherent}.

Decomposition techniques are nowadays an almost ubiquitous analysis method in fluid mechanics, being a preliminary step towards flow modelling, control and optimization. Data-driven decomposition techniques, in particular, have gained traction due to their flexibility and the abundance of large flow datasets generated by modern experimental and numerical methods. Among them, \acl{POD} \citep[\acs{POD}\acused{POD},][]{lumley1967structure,lumey1970stochastic,berkooz1993proper} occupies a prominent role in terms of usage. The method is popular in other disciplines mainly with the names principal component analysis (PCA), empirical orthogonal function (EOF) analysis and Karhunen–Loève
decomposition.  More specifically, the literature has been mostly dominated by a specific version of the \ac{POD}, nowadays mostly referred as \emph{standard}, \emph{space-only} or \emph{snapshot} \citep{sirovich1987turbulence,aubry1991hidden,aubry1991spatiotemporal}, since it does not impose any requirement on the data fed to it -- especially in terms of temporal resolution. The space-only \ac{POD} identifies a set of real-valued orthonormal eigenfunctions, i.e. modes, which provide the most compact representation, in a least-square sense, of a set of observations of the flow field, which typically represent snapshots of the field at different observation times. 
These modes, even if uncorrelated, may share a mutual spatiotemporal dependency due to the inability of real-valued functions to embed phase information. Advecting structures, acting as travelling wavepackets, are typically split in pairs of neighbouring real-valued POD modes, sharing a similar frequency content (both in time and space) but in phase quadrature. In periodic flows this dependency can be easily spotted by reconstructing the temporal modes in so-called Lissajous figures, which reveal the phase-relation in the corresponding shape of the resulting cyclograms \citep[see e.g. ][]{ben2004reconstruction,raiola2016wake}. Simple oscillator models can be constructed by summing these pairs as real and imaginary part of a complex-valued mode \citep{nair2018networked}. Since spatiotemporal coherence is not enforced in the original \ac{POD} problem, however, modes in more complex flow field may fail in providing a clear description of travelling features.

An alternative decomposition, the \ac{DMD}, was developed by \cite{schmid2010dynamic} to specifically identify coherent flow structures describing the flow dynamics, i.e. the evolution of the flow through time. The work was motivated by loss of dynamical information produced by the space-only POD, specifically due to the averaging process used to obtain the spatial correlation tensor. \ac{DMD} is based on the eigendecomposition of a best-fit linear operator that
approximates the dynamics present in the data. Its eigenvalues can be interpreted as the temporal evolution embedded in the linear dynamics tensor, i.e. complex-valued frequencies of an exponential function, making \ac{DMD} a data-driven equivalent to linear stability analysis. The capability of \ac{DMD} to decompose the dynamics of the flow, roots in the Koopman theory \citep{koopman1931hamiltonian}, consisting in approximating non-linear dynamical system with a much larger (typically infinite dimensional) linear one. A deeper review of the connection between Koopman analysis and \ac{DMD} can be found in \cite{budivsic2012applied,mezic2013analysis,tu2013dynamic}.

For statistically stationary flows, spatiotemporal coherent modes can be extracted by the \acl{SPOD} \citep[\acs{SPOD}\acused{SPOD}, ][]{lumey1970stochastic,towne2018spectral}, which restates the original POD problem in the frequency space rather than in temporal coordinates. Differently from space-only POD, which is based on the decomposition of the two-point correlation matrix, the \ac{SPOD} extracts the modes from the cross-spectral density matrix for each specific temporal frequency. This results in a set of orthogonal complex-valued spatial modes each linked to an harmonic temporal evolution. As such, \ac{SPOD} provides a decomposition, which is spatiotemporally coherent. In particular, thanks to the translation properties of the Fourier transform, \ac{SPOD} has became one of the standard techniques to extract advecting patterns, as for example with wavepackets in turbulent jets \citep{cavalieri2019wave}. One of the main drawbacks of the technique, however, stems from the hypothesis of statistical stationarity and the use of Fourier transform: the temporal evolution is fixed to follow a pure harmonical relation having infinite temporal support, i.e. extending up to infinity, thus losing any instantaneous character of the temporal evolution. Additionally, \ac{SPOD} imposes spectral purity on the temporal coefficients of the modes, thus separating in different modes coherent structures with multiple frequency content. This is especially problematic in highly turbulent flows, where coherent structures can be characterized by strong intermittency and variable-frequency due to phase jitter or frequency modulations. In these cases, a single physical feature is, therefore, likely to be splitted among multiple \ac{SPOD} modes.

Alternative extensions of the \ac{POD} have been proposed by \cite{sieber2016spectral} and \cite{mendez2019multi} blending \ac{DFT} and \ac{POD}. Sieber's spectral proper orthogonal decomposition (not to be confused by the one proposed in the original work by \citealp{lumey1970stochastic} and re-popularized by \citealp{towne2018spectral}) applies a low-pass filter along the diagonals of the temporal correlation matrix before proceeding to compute the \ac{POD} according to the snapshot method. This filter forces the matrix to assume a form closer to a Toeplitz circulant matrix, whose eigenfunctions are Fourier modes, according to the Szegö theorem \citep{grenander1958toeplitz}. Depending on the filter size, the method can blend between pure \ac{DFT} and pure POD. The \ac{mPOD} introduced by \cite{mendez2019multi} combines the idea of windowed \ac{SPOD} (borrowed from \citealp{sieber2016spectral}) and the \ac{SPOD} in \cite{towne2018spectral}. Instead of filtering the correlation  as in \cite{sieber2016spectral}, the \ac{mPOD} decomposes it into the contributions of different scales using the multi-resolution architecture. Instead of computing various eigenbases from the spectra of different portions of the data, as in \cite{towne2018spectral}, the \ac{mPOD} computes eigenbases on the correlation matrix of different scales. While these techniques can provide a decomposition with a more complex spectral behavior, however, they also introduce user-dependent parameters in the process, thus effectively not delivering an univocal decomposition.

\cite{kriegseis2021hilbert} introduced in the field of fluid mechanics an extension of the \ac{POD}, under the name of \ac{HPOD}, which can automatically deliver complex-valued modes satisfying the conditions to build oscillator models. The method roots in the analytic signal concept to obtain a complex-valued extension of flow-field time series prior to performing the space-only \ac{POD}. The complex-valued extension is obtained through the Hilbert transform \citep{hahn1996hilbert}, adding to the original time-series an imaginary part shifted by $\pi/2$ in time, independently on the frequency content, to build the analytic signal of the field time series. The \ac{HPOD} on time series, however, requires time-resolved field data, converging to standard \ac{POD} when undersampled data are used \citep[see e.g.][]{kriegseis2021hilbert}. 
The use of the Hilbert transform and analytic signal in the temporal direction for fluid mechanics applications is indeed not novel: examples of it can be found for measuring the intermittency in turbulence phenomena \citep{sreenivasan1985fine} or to identify envelope modulation of fine turbulence scales \citep{mathis2009large}.
While the name \ac{HPOD} has been used for the first application to flow field analysis, the technique has been originally introduced with different names in several fields of study. The first application of the technique, under the name of \ac{COEF},
targeted the identification of coherent propagating features in gridded wind data for meteorological applications, where the short-term irregular nature of some of the features made common spectral analysis inappropriate \citep{barnett1983interaction1,barnett1983interaction2,barnett1983interaction3}. \cite{horel1984complex} demonstrated the soundness of the technique, renamed \ac{CPC} analysis, to extract travelling and standing waves from geophysical data. \cite{pfeffer1990study} applied \ac{CPC} analysis in a thermally-driven, rotating annulus fluid flow with sinusoidal bottom topography for meteorological applications. The limitation of the \ac{COEF} in isolating widebanded propagating waves has been studied by \cite{merrifield1990detecting}. \ac{CPC} analysis has been used to analyze the temporal variation of sea surface temperature \citep{alessio1999space}.
\cite{feeny2008complex} introduced the technique under the name \ac{COD} for identifying wave motion in vibration analysis, also addressing an index to classify standing and traveling waves. \ac{COD} has been used for the kinematic study of flexible structures in fluid-structure interaction applications \citep{yeaton2020undulation,leroy2022tapered}. \ac{CPC} analysis has been applied to the temporal variation of the Antarctic ice sheet \citep{zhan2021complex} as well as to MRI data to study the organization of the human brain \citep{bolt2022parsimonious}.

Other strategies to identify travelling features involve a change of the reference frame in which space-only \ac{POD} is performed. As pointed out by \cite{brunton2022data}, travelling features introduce translation symmetries in the fields, which disrupts the spatial alignment of coherent features between snapshots and therefore corrupts the correlation matrix, and provides a decomposition which splits features over multiple modes. To provide a correct application of the space-only \ac{POD}, translation invariance must be taken into account. For wave-like patterns, this involves accounting for the phase velocity of the wave. Procedures like \emph{centring} \citep{glavaski1998model}, i.e. shifting the wave so that centres align between frames, or \emph{template fitting} \citep{rowley2000reconstruction,rowley2003reduction}, i.e. shifting the data so to maximize the correlation with a preselected template, allow to account for a single advecting velocity. Multiple advecting velocities can be accounted through the \emph{shifted} \ac{POD} \citep{reiss2018shifted}, where modes advecting at different velocities are separated into different reference frames through an iterative procedure. This class of strategies to identify advecting features relies on the estimation of the advection velocity and using it to correct the translation symmetry explicitly.

Following a similar logic of correcting for the translation symmetry through a reference frame transformation, \cite{sesterhenn2019characteristic} introduced the concept of performing the decomposition in a spatiotemporal direction characteristic of the travelling structure. This transformation is introduced by performing a rotation in the spatial and temporal dimensions of the data, identifying the rotation -- and hence the advection velocity -- maximizing the difference between the first two singular values. Performing the modal analysis in a direction different from the temporal one has already been exploited several times for advecting features. \cite{schmid2010dynamic} considers also performing the DMD in the spatial direction rather than the temporal one to provide a data-driven \emph{spatial stability analysis}. This same idea was later exploited by \cite{ek2022permuted} to produce the \emph{permuted} \ac{POD}. By reorienting the data matrix along the advection direction through permutation prior to performing standard POD, they produced modes, which are orthogonal in time and in the remaining spatial directions. Similar approaches can also be identified in non-data-driven decompositions. For example, \cite{encinar2020momentum} applied complex wavelet analysis in the advection direction to identify wavetrains in turbulent channel flows.

All the decompositions targeting travelling features reported so far rely on the availability of temporally resolved data in order to enforce the spatiotemporal coherence of the modes. When no temporal resolution is available, only standard POD can be applied, trying to reconstruct spatiotemporal coherent features pairing real-valued modes. The present work elaborates on the capabilities of \ac{HPOD} to provide a decomposition targeting travelling wave flow features and introduce a \emph{space-only} version of the technique, which does not require temporal resolution of the data. This \emph{space-only} \ac{HPOD} is based on computing the analytic signal from the Hilbert transform in the advection direction, instead than in the temporal direction of the dataset, as in the \emph{conventional} \ac{HPOD}. This means that the imaginary part of the analytic signal is shifted by $\pi/2$ along the advection direction with respect to the original real-valued signal.
Therefore, the space-only \ac{HPOD} can target advecting wavepackets by leveraging the similarity between time and advection direction in advecting flows.  In this sense, the approach has some connection with the permuted \ac{POD} by \cite{ek2022permuted}. 

Section \ref{sec:math} presents a general framework for both the conventional and space-only \ac{HPOD}, introducing the properties of the decomposition, providing the interpretation of the complex-valued wavepacket functions as analytic signals and demonstrating that both conventional and space-only \ac{HPOD} deliver eigenfunctions, which are analytic signals in time for advecting flows.
Section \ref{sec:testcases} provides 3 different test cases to demonstrate the properties of the \ac{HPOD} and highlight its capabilities. The first dataset is a \ac{DNS} of a laminar wake of a cylinder, with a compact and spectrally pure shedding signature, used to demonstrate the properties of both the conventional and space-only \ac{HPOD}. The second test case is the \ac{LES} of a turbulent jet, featuring more complex wavepackets, which is used to show the capabilities of \ac{HPOD} to capture amplitude-modulated and widebanded travelling waves. The third dataset are velocity fields of a turbulent jet measured via snapshot \ac{PIV}, which is used to show the robustness of the space-only \ac{HPOD} in real experimental scenarios. Finally, Section \ref{sec:concl} draws the conclusion on both \ac{HPOD}s versions, suggesting their possible role in the toolbox of researchers in the field of fluid mechanics.

\section{Mathematical background}\label{sec:math}

\subsection{Standard Proper Orthogonal Decomposition}\label{sec:POD}
The \emph{standard} (also referred to as \emph{space-only}) \acf{POD} aims at determining the set of functions $\psi_j (t)$ and $\bm{\phi}_j (\bm{x})$, which best approximates a field $\bm{u}(\bm{x},t)$. Here $\bm{x}=(x,y,z)$ and $t$ represent the spatial and temporal coordinates, respectively.  In the following, the bi-orthogonal formulation introduced by \cite{aubry1991spatiotemporal} will be followed and recapped to describe the mathematical framework of this method. The decomposition of $\bm{u}$ aims at identifying a set of temporal orthogonal modes $\psi_j$ and spatial orthogonal modes $\bm{\phi}_j$ so that
\begin{equation}
\bm{u}(\bm{x},t)=\sum_{j} \psi_j (t) \sigma_j \bm{\phi}_j(\bm{x})\,,
 \label{eq:POD}
\end{equation}
where $\sigma_j$ is the singular value associated to each $\psi_j$ and $\bm{\phi}_j$. Notice that $\sigma_j$ is always a non-negative real. The functions $\psi_j$ and $\bm{\phi}_j$, instead, will be either real- or complex-valued  -- depending on whether the original field $\bm{u}$ is real- or complex-valued. Both the functions $\bm{\phi}_j$ and $\psi_j$ form an orthogonal basis, i.e.
\begin{equation}
\begin{split}
&\int_\Omega \bm{\phi}_j(\bm{x})\bm{\phi}_k^*(\bm{x})\dx\bm{x}=\delta_{jk}\,,\\
&\int_T \psi_j(t)\psi_k^*(t)\dx t=\delta_{jk}\,,
\end{split}
\end{equation}
where $\Omega$ and $T$ refer to the spatial and temporal domain over which the field is considered.

The decomposition in Eq. \eref{eq:POD} is solved by identifying the set of orthogonal functions $\bm{\phi}_j$ solving the two equivalent minimization or maximization problems  
\begin{equation}\label{eq:minspace}
\min_{\bm{\phi}} \left\langle\left\| \bm{u}-\frac{\int_\Omega \bm{u}(\bm{x},t)^* \bm{\phi}_j(\bm{x}) \dx\bm{x}}{\| \bm{\phi}_j(\bm{x})\|_x^2} \bm{\phi}_j(\bm{x}) \right\|_x^2\right\rangle
\Leftrightarrow
\max_{\bm\phi} \frac{\left\langle|\int_\Omega \bm{u}(\bm{x},t)^* \bm{\phi}_j(\bm{x}) \dx\bm{x}|^2 \right\rangle}{\| \bm{\phi}_j(\bm{x})\|_x^2}\,,
\end{equation}
where $^*$ indicates the complex conjugate, $\langle \cdot \rangle$ indicates the ensemble average and $\|\cdot\|_x^2$ indicates the spatial norm $\|f\|_x^2=\int_\Omega  f^*(\bm{x}) f(\bm{x}) \dx\bm{x}$.
Similarly, the decomposition can be restated as the identification of the set of functions $\psi_j$ solving the minimization or maximization problems 
\begin{equation}\label{eq:mintime}
\min_\psi \left\langle\left\| \bm{u}-\frac{\int_T \bm{u}(\bm{x},t)^* \psi_j(t) \dx t}{\| \psi_j(t)\|_t^2} \psi_j(t) \right\|_t^2\right\rangle
\Leftrightarrow
\max_\psi \frac{\left\langle|\int_T \bm{u}(\bm{x},t)^* \psi_j(t) \dx t|^2 \right\rangle}{\| \psi_j(t)\|_t^2}\,,
\end{equation}
where $\|\cdot\|_t^2$ indicates the temporal norm $\|f\|_t^2=\int_T  f^*(t) f(t) \dx t$.

The ensemble average $\langle \cdot \rangle$ in Eqs. \eref{eq:minspace} and \eref{eq:mintime} is typically defined either as the average of the samples in time (for the spatial norm problem) or space (for the time norm problem) when considering standard \ac{POD}. The solution of the problem in Eq. \eref{eq:minspace} is given by the Fredholm integral equation
\begin{equation}\label{eq:spaceeigprob}
    \int_\Omega C(\bm{x},\bm{x}',t=t') \bm{\phi}_j(\bm{x}') \dx\bm{x}'=\sigma_j^2 \bm{\phi}_j(\bm{x})\,,
\end{equation}
which represents the eigenvalue problem for the two-(space)-point correlation function $C(\bm{x},\bm{x}',t=t')$
\begin{equation}
    C(\bm{x},\bm{x}',t=t')=\langle \bm{u}(\bm{x},t) \bm{u}^*(\bm{x}',t') \rangle=\int_T \bm{u}(\bm{x},t) \bm{u}^*(\bm{x}',t) \dx t\,.
\end{equation}
Similarly, the  solution for the problem in Eq. \eref{eq:mintime} is provided by the eigenvalue problem
\begin{equation}\label{eq:timeeigprob}
\int_T C(\bm{x}=\bm{x}',t,t')\psi_j(t')\dx t'=\sigma_j^2\psi_j(t)\,,
\end{equation}
where $C(\bm{x}=\bm{x}',t,t')$ is the two-(time)-instant correlation function 
\begin{equation}
C(\bm{x}=\bm{x}',t,t')=\langle \bm{u}(\bm{x},t) \bm{u}^*(\bm{x}',t') \rangle=\int_\Omega \bm{u}(\bm{x},t) \bm{u}^*(\bm{x},t')\dx\bm{x}\,.
\end{equation}

Following the bi-orthogonal approach, the orthogonal temporal and spatial modes are obtained as the eigenfunctions of the two symmetric eigenvalue problems in space (Eq. \ref{eq:spaceeigprob}) and time (Eq. \ref{eq:timeeigprob}). These modes are not independent one from the other, sharing instead a one-to-one correspondence between spatial and temporal parts, as shown by Eq. \eref{eq:POD}. In the following, the denomination proposed by \cite{aubry1991hidden} will be adopted, referring to each term of the summation in Eq. \eref{eq:POD} -- i.e. a pair of temporal and spatial modes ($\psi_j$ and $\bm{\phi}_j$) and their associated eigenvalue ($\sigma_j$) -- as \emph{structure}.

The adjective \emph{space-only}, oftentimes used to refer to this decomposition, reflects the fact that spatiotemporal coherency is not considered, i.e. the problem either solves the eigenvalue problem for zero-time-lag spatial correlation matrix or the zero-space-shift temporal correlation matrix. In this sense, the decomposition is not guaranteed to effectively track coherent flow features, which evolve through time and space. Also, the nomenclature used here suggests that the two-point and the two-instant correlation matrices are special cases of the space-time correlation tensor
\begin{equation}
C(\bm{x},\bm{x}',t,t')=\langle \bm{u}(\bm{x},t) \bm{u}^*(\bm{x}',t') \rangle\,,
\end{equation}
whose eigenvalue problem produces the \emph{space-time} POD.

\subsection{Wave-packet interpretation of coherent structures} \label{sec:wavepacket}
The aim of this work is to identify coherent features evolving spatiotemporally in advective flows. These features can be interpreted as wave packets travelling in the flow direction. A single wave packet can be represented (here in 1 dimension for ease of notation) as a sinusoidal function,
\begin{equation}
w_j(x,t)=A_j(x,t) \cos(k_j x-\omega_j t)=A_j(x,t) \cos[k_j( x-c_j t)]\,,
\end{equation}
where $A_j$ represents its instantaneous and local amplitude, $k_j$ its spatial angular wavenumber, $\omega_j$ its temporal angular frequency and $c_j=\omega_j/k_j$ its wave velocity. This wave packet can be considered as the real part of a complex-valued wave, constructed by adding a companion imaginary sinusoidal function shifted $\pi/2$ with respect to the original wave,
\begin{equation}
\begin{split}
w_j(x,t)&=A_j(x,t) \text{Re}[\cos(k_j x-\omega_j t)+i \sin(k_j x-\omega_j t)]\\ &=A_j(x,t) \text{Re}[e^{ i (k_j x-\omega_j t)}]\,,
\end{split}
\end{equation}
where \text{Re} indicates that the real part of the function should be taken, $i$ indicating the imaginary unit and where the equality with the exponential function follows from Euler's formula. It is worth noticing that the sinusoidal functions can be further decomposed using the trigonometric formulas
\begin{equation}\label{Eq:sinusoida}
\begin{split}
e
^{ i(k_j x-\omega_j t)}=&\left[\cos(k_j x) \cos(\omega_j t) + \sin(k_j x) \sin(\omega_j t)\right] \\&+ i \left[\sin(k_j x) \cos(\omega_j t) - \cos(k_j x) \sin(\omega_j t)\right]\,.
\end{split}
\end{equation}

The imaginary part of Eq. \eref{Eq:sinusoida} can be interpreted both as a $\pi/2$ phase shift operated on the temporal sinusoidal functions, i.e.
\begin{equation}
\begin{split}
e^{ i(k_j x-\omega_j t)}=&
\cos(k_j x) [\cos(\omega_j t) - i  \sin(\omega_j t)] + \sin(k_j x) [\sin(\omega_j t) + i \cos(\omega_j t)]\\
=&F(x,t)+i\,F\left(x,t+\frac{1}{\omega_j}\frac{\pi}{2}\right)\,,
\end{split}
\end{equation}
or as a phase shift on the spatial ones, i.e.
\begin{equation}
\begin{split}
e^{ i\,(k_j x-\omega_j t)}=&
\cos(\omega_j t) [\cos(k_j x) + i\, \sin(k_j x) ] + \sin(\omega_j t) [\sin(k_j x) - i\, \cos(k_j x)]\\
=&F(x,t)+i\,F\left(x+\frac{1}{k_j}\frac{\pi}{2},t\right).
\end{split}
\end{equation}

In order to extract the wavepackets from a given flow field, including all their phase and amplitude information, the data needs to be expanded to the complex domain. This operation is termed complexification and is described in the following section.

\subsection{Complexification of the flow field} \label{sec:complexification}
Several techniques can be applied to obtain a complex-valued version of the flow field. Spectral POD \citep[\acs{SPOD},][]{towne2018spectral, schmidt2020guide}, for example, performs a Fourier decomposition in time over the flow field sequence. According to this complexification, the \ac{SPOD} is the result of an eigenvalue problem on the cross-spectral density tensor. However, this means that the sequence, and therefore the resulting modes, is expressed in the frequency space.

An alternative operation, popular in signal analysis, is the computation of the analytic signal, which consist in removing the redundant negative frequency components from the original real signal. This operation recovers a complex-valued signal whose real part is identical to the original signal, and which has a single-sided spectrum that preserves the original signal frequency content. 
The analytic signal is typically constructed by adding to the original real-valued signal an imaginary signal obtained from the Hilbert transform of the original signal. For one-dimensional  real-valued temporal signal $s(t)$, the analytic signal is then expressed as
\begin{equation}
\tilde{s}(t)=s(t)+i\,\mathcal{H}_t[s(t)]\,,
\end{equation}
where $i$ is the imaginary unit and $\mathcal{H}_t$ is the Hilbert transform operator \citep{hahn1996hilbert} applied in time and defined as the Cauchy principal value of the convolution of $s(t)$ with the signal $1/\pi t$ according to
\begin{equation}
\mathcal{H}_t[s(t)]=\lim_{e\rightarrow0}\int_{|\tau-t|>e} \frac{s(\tau)}{\pi(t-\tau)}\dx\tau=\frac{1}{\pi t} \ast s(t)\,,
\end{equation}
where $\ast$ indicates the convolution operation. The analytic signal of $s(t)$ can therefore be written using the convolution notation as
\begin{equation}
\tilde{s}(t)=\left(\delta(t)+\frac{i}{\pi t}\right) \ast s(t)\,,
\end{equation}
with $\delta(t)$ being the Dirac delta function.

The Hilbert transform can be interpreted in the Fourier space as a $\pi/2$ shift, either in advance or delay depending on the sign of the frequency component. If $\mathcal{F}$ is used to indicate the Fourier transform \citep{hahn1996hilbert}:
\begin{equation}\label{eq:hilbertfourier}
    \mathcal{F}\left\{\mathcal{H}[s]\right\}(f)= -\text{sign}(f) e^{-i\frac{\pi}{2}}  \mathcal{F}[s](f)=-i\,\text{sign}(f) \mathcal{F}[s](f)
\end{equation}
which leads to the following expression for the analytic signal: 
\begin{equation}
    \tilde{s}(t)=\mathcal{F}^{-1}\left\{[1+i\,\text{sign}(f)] \mathcal{F}[s(t)]\right\}
\end{equation}
This expression shows that the analytic signal shares the same frequency content of the original signal for positive frequencies, while the content for negative frequencies is discarded. Additionally, this expression provides the operative definition used in this work for the computation of the analytic signal from a discrete signal: the \ac{FFT} of the signal is first computed, then negative frequency terms are set to zero while positive frequency terms are doubled, and finally the Inverse \ac{FFT} is applied. It is worth highlighting at this point that, since \ac{FFT} is involved, this algorithm does not recover the exact analytic signal for finite sequences due to the edge effects introduced by the Fourier transform. These edge effects are typically treated by removing the corrupted part of the signal before further analyses are carried out. The implications of these errors on the proposed technique will be exemplified in Section \ref{sec:cyltestcase}. Other techniques to correct edge effect might involve padding or windowing of the original signal, which, however, will not be explored in this work.

The analytic signal obtained for an oscillatory signal can be interpreted as a generalized phasor of the original signal, according to the polar representation for complex values
\begin{equation}
\tilde{s}(t)=|s(t)| e^{\varphi(t)}\,,
\end{equation}
which contains information on its instantaneous amplitude $|s(t)|$ and phase $\varphi(t)$. This concept is at the basis of the Hilbert spectral analysis, which allows to extract the instantaneous frequency of a signal as
\begin{equation}
    f(t)=\frac{1}{2\pi}\frac{\dx\varphi(t)}{\dx t}\,.
\end{equation}

The Hilbert transform in time can be extended to a generic real-valued velocity field $\mathbf{u}(x,y,z,t)$ to obtain its complex-valued analytic representation
\begin{equation}
\tilde{\mathbf{u}}(x,y,z,t)=\mathbf{u}(x,y,z,t)+i\,\mathcal{H}_t[\mathbf{u}(x,y,z,t)]
\end{equation}
with
\begin{equation}
    \mathcal{H}_t[\mathbf{u}(x,y,z,t)]=\lim_{e\rightarrow0}\frac{1}{\pi}\int_{|\tau-t|>e} \frac{\mathbf{u}(x,y,z,\tau)}{t-\tau}\dx\tau\,.
\end{equation}

Alternatively, the Hilbert transform $\mathcal{H}$ can be defined along one of the spatial directions. Assuming an advection-dominated flow, with $x$ being the direction in which the flow structures are advected, this operation would provide an analytic representation whose imaginary part is shifted by $\pi/2$ in the $x$ direction. This would result in the analytic representation
\begin{equation}
\tilde{\mathbf{u}}(x,y,z,t)=\mathbf{u}(x,y,z,t)+i\,\mathcal{H}_x[\mathbf{u}(x,y,z,t)]
\end{equation}
with $\mathcal{H}_x$ being the Hilbert transform in the advection direction, i.e.
\begin{equation}
\mathcal{H}_x[\mathbf{u}(x,y,z,t)]=\lim_{e\rightarrow0}\frac{1}{\pi}\int_{|\xi-x|>e} \frac{u(\xi,y,z,t)}{x-\xi}\dx\xi\,.
\end{equation}

While the two analytic representations of a flow field might appear contrasting, they can be reconciled if the flow field is assumed to behave as a wave packet. In fact, following Section \ref{sec:wavepacket}, it can be shown that
\begin{equation}
\begin{split}
    e^{i(k_j x -\omega_j t)}=&f(x,t)+i\,f\left(x,t+\frac{1}{\omega_j}\frac{\pi}{2}\right)=f(x,t)+i\, \mathcal{H}_t[f(x,t)]\\=&f(x,t)+i\,f\left(x+\frac{1}{k_j}\frac{\pi}{2},t\right)=f(x,t)+i\, \mathcal{H}_x[f(x,t)]\,.
\end{split}
\end{equation}
Since the Hilbert transform is a linear operator, the same can be applied to a field composed of a sum of wave packets. Based on this concept, the proposed complexification of the flow field allows for recognizing complex-valued wavepackets. In the following, a complex-valued version of the \ac{POD} is introduced, which allows the distilling of wave-packet-like features from the complexified flow field.

\subsection{Hilbert Proper Orthogonal Decomposition in time and space}
Once the field has been extended to complex values, applying the standard POD as introduced in Section \ref{sec:POD} is possible. The decomposition of $\tilde{\bm{u}}$ identifies a set of complex-valued temporal ($\psi_j$) and spatial ($\bm{\phi}_j$) orthogonal modes, i.e.
\begin{equation}
\tilde{\bm{u}}(\bm{x},t)=\sum_{j} \psi_j (t) \sigma_j \bm{\phi}_j(\bm{x})\,,
 \label{eq:HPOD}
\end{equation}

 resulting from the solution of the eigenvalue problem for the two-(space)-point correlation function of the analytic field representations
\begin{equation}\label{eq:HPOD_spacecorr}
\begin{split}
&\int_\Omega \tilde{C}(\bm{x},\bm{x}',t=t') \bm{\phi}_j^*(\bm{x}') \dx\bm{x}'=\sigma_j^2 \bm{\phi}_j(\bm{x})\,,\\
&\tilde{C}(\bm{x},\bm{x}',t=t')=\int_T \tilde{\bm{u}}(\bm{x},t) \tilde{\bm{u}}^*(\bm{x}',t) \dx t\,,
\end{split}
\end{equation}
or, similarly, of the eigenvalue problem for the two-(time)-instant correlation function
\begin{equation}\label{eq:HPOD_timecorr}
\begin{split}
&\int_T \tilde{C}(\bm{x}=\bm{x'},t,t')\psi_j^*(t')\dx t'=\sigma_j^2\psi_j(t)\,,\\
&\tilde{C}(\bm{x}=\bm{x'},t,t')=\int_\Omega \tilde{\bm{u}}(\bm{x},t) \tilde{\bm{u}}^*(\bm{x},t')\dx\bm{x}\,.
\end{split}
\end{equation}

Since the correlation matrix is obtained from analytic signals, the eigenfunctions obtained from this approach would be analytic signals as well, i.e. they would have zero negative frequencies content. A proof of this is provided in the following section. Therefore, the HPOD eigenfunctions would allow the extraction of instantaneous/local frequency, phase and amplitude of each eigenfunction, similarly to what would be possible through an Hilbert spectral analysis.

\subsection{The analytic signal nature of HPOD eigenfunctions}
It is possible to prove that the HPOD problem delivers eigenfunctions, which are analytic signals, i.e. that their spectral content is limited to positive frequency only. Consider the analytic signal of the temporal correlation function $\tilde{C}$. Its relation with the cross-spectral density function is given by 
\begin{equation}\label{eq:crossspectraltime}
        \tilde{C}(x,x',t,t')=\tilde{C} (x,x',t,\tau)=\int_{-\infty}^\infty (1+\text{sign}(\omega))  S(x,x',t,\omega )  e^{i \omega \tau}   \dx\omega
\end{equation}
where $S(x,x',t,\omega)$ is the cross-spectral density function at an instant $t$ (in the time-frequency sense), $\omega$ represent the temporal angular frequency, and $\tau=t-t'$ \citep[see e.g.][]{hahn1996hilbert}. Notice that for ease of notation a single spatial coordinate, i.e. the advection direction, is considered in the following, but results will be equally valid for multiple spatial coordinates.
In Eq. \eref{eq:crossspectraltime} the Fourier transform and the Hilbert transform are taken in the temporal direction. The temporal POD problem over an infinite temporal interval is given by
\begin{equation}
\int_{-\infty}^\infty \tilde{C} (x,x',t,\tau )  \bm{w}(x',t' )  \dx t'=\sigma^2 \bm{w}(x,t) \,.
\end{equation}
Substituting the Fourier transform of $\tilde{C}$ yields
\begin{equation}
\begin{split}
&\int_{-\infty}^\infty \left[\int_{-\infty}^\infty(1+\text{sign}(\omega ))  S(x,x',t,\omega )  e^{i \omega\tau }  \dx \omega \right]   \bm{w}(x',t' )  \dx t'=\sigma^2 \bm{w}(x,t) \,,\\
&\int_{-\infty}^\infty (1+\text{sign}(\omega))  S(x,x',t,\omega )  e^{i\omega t }  \left[ \int_{-\infty}^\infty \bm{w}(x',t' ) e^{-i\omega t' }    \dx t'\right]  \dx \omega =\sigma^2 \bm{w}(x,t) \,,\\
&\int_{-\infty}^\infty (1+\text{sign}(\omega))  S(x,x',t,\omega)  e^{i\omega t }  \mathcal{F}[{\bm{w}}](x',\omega )  \dx \omega =\sigma^2 \bm{w}(x,t) \,.\\
\end{split}\label{eq:2.32}
\end{equation}

Assuming spectrally-pure eigenfunctions (or equivalently a single component of the Fourier expansion of said eigenfunctions), the eigenfunction and its Fourier transform are
\begin{equation}
\begin{split}
&\bm{w}(x,t)=\bm{q}(x,\omega' ) e^{i\omega' t } \,,\\
&\mathcal{F}[{\bm{w}}](x,\omega) =\bm{q}(x,\omega' )\delta (\omega-\omega') \,.
\end{split}\label{eq:2.33}
\end{equation}

Substituting \eref{eq:2.33} into the eigenproblem of \eref{eq:2.32} then leads to
\begin{equation}
\int_{-\infty}^\infty (1+\text{sign}(\omega))  S(x,x',t,\omega)  e^{i\omega t }  \bm{q}(x',\omega' )\delta (\omega-\omega')  \dx \omega =\sigma^2 \bm{q}(x,\omega' ) e^{i\omega' t }\,.\label{eq:2.34}
\end{equation}

Now it is possible to divide the eigenproblem \eref{eq:2.34} in two, depending of the sign of $\omega$
\begin{equation}\label{eq:SPOD}
\begin{dcases}
\text{if } \omega<0: & 0=\sigma^2 \bm{q}(x,\omega )\\
 \text{if } \omega>0: & S(x,x',t,\omega )  \bm{q}(x',\omega ) =\frac{\sigma^2}{2} \bm{q}(x,\omega ) \,.
\end{dcases} 
\end{equation}

The previous equation shows that the eigenproblem solved for the HPOD in time produces temporal eigenfunctions with only positive frequencies, i.e. eigenfunctions, which are analytic signals.

Considering instead the analytic signal of the spatial correlation function  
\begin{equation}\label{eq:crossspectralspace}
        \tilde{C}(x,x',t,t')=\tilde{C} (x,\xi, t,t')=\int_{-\infty}^\infty (1+\text{sign}(k))  S(x,k, t,t')  e^{i k \xi}   \dx k \,,
\end{equation}
where $S(x,k, t,t')$ is the cross-spectral density function at a given position $x$, $k$ represent the spatial angular wavenumber, and $\xi=x-x'$.
In Eq. \eref{eq:crossspectralspace} the Fourier transform and the Hilbert transform are taken in the spatial direction. The spatial POD problem over an infinite spatial interval is given by
\begin{equation}
\int_{-\infty}^\infty \tilde{C} (x,\xi, t,t')  \bm{w}(x',t' )  \dx x'=\sigma^2 \bm{w}(x,t) \,.
\end{equation}
As above, substituting the Fourier transform of $\tilde{C}$ yields
\begin{equation}
\begin{split}
&\int_{-\infty}^\infty \left[\int_{-\infty}^\infty(1+\text{sign}(k ))  S(x,k, t,t')  e^{i k\xi }  \dx k \right]   \bm{w}(x',t' )  \dx x'=\sigma^2 \bm{w}(x,t) \,,\\
&\int_{-\infty}^\infty (1+\text{sign}(k))  S(x,k, t,t')  e^{ik x }  \left[ \int_{-\infty}^\infty \bm{w}(x',t' ) e^{-ik x' }    \dx x'\right]  \dx k =\sigma^2 \bm{w}(x,t) \,,\\
&\int_{-\infty}^\infty (1+\text{sign}(k))  S(x,k, t,t')  e^{ik x }  \mathcal{F}[{\bm{w}}](k,t' )  \dx k =\sigma^2 \bm{w}(x,t) \,.
\end{split}\label{eq:2.38}
\end{equation}
The eigenfunction and its Fourier transform can be assumed to be
\begin{equation}
\begin{split}
&\bm{w}(x,t)=\bm{q}(k',t ) e^{ik' x } \,,\\
&\mathcal{F}[{\bm{w}}](k,t) =\bm{q}(k',t )\delta (k-k')\,,
\end{split}
\end{equation}
which substituted into the eigenproblem of Eq. \eref{eq:2.38} leads to
\begin{equation}
\int_{-\infty}^\infty (1+\text{sign}(k))  S(x,k,t,t')  e^{ik x }  \bm{q}(k',t')\delta (k-k')  \dx k =\sigma^2 \bm{q}(k',t ) e^{ik' x }\,.
\end{equation}
Again, it is possible to divide the eigenproblem in two, depending of the sign of $k$ according to
\begin{equation}\label{eq:SPOD}
\begin{dcases}
\text{if } k<0: & 0=\sigma^2 \bm{q}(k,t )\\
 \text{if } k>0: & S(x,k,t,t' )  \bm{q}(k,t') =\frac{\sigma^2}{2} \bm{q}(k,t) \,,
\end{dcases} 
\end{equation}
which proves that the eigenproblem solved for the HPOD in space produces spatial eigenfunctions that are analytic signals.

\subsection{Relation between the HPOD and 
the SPOD}

In the previous section, it has been demonstrated that the conventional \ac{HPOD} delivers temporal modes, which are analytic signals; conversely, space-only \ac{HPOD} has demonstrated to provide spatial modes, which are analytic signals. It is possible to show that for structures capturing advecting features both temporal and spatial modes need to be analytic under the assumption that the spatiotemporal evolution is dictated by a phase velocity relationship. This translates in the two \ac{HPOD} approaches being equivalent for a travelling wavepacket. In the following, this property is demonstrated in the setting of stationary and homogeneous fields, both for ease of development as well to highlight the linking points with another popular decomposition technique, the \ac{SPOD}.

For a stationary and homogeneous (in the advection direction $x$) field, the analytic signal of the time-space correlation tensor $C(x,x',t,t')$ is given by
\begin{equation}\label{eq:crossspectral}
        \tilde{C} (x,x',t,t')=\tilde{C} (\xi,\tau)=\int_{-\infty}^\infty (1+\text{sign}(k))  S(k,\omega)  e^{i (k \xi-\omega \tau)}   \dx k\,,
\end{equation}
where $S(k,\omega)$ is the cross-spectral density tensor, $k$ and $\omega$ respectively represent the angular wavenumber and frequency in space and time, 
and 
\begin{equation}
\left\{\begin{split}
& \xi=x-x'\\
& \tau=t-t'\\
\end{split}\right.
\end{equation}

In Eq. \eref{eq:crossspectral} the Fourier transform and the Hilbert transform are taken in space. However a similar final result would be obtained if the temporal direction were taken. 

The spatial HPOD problem over an infinite spatial interval is given by
\begin{equation}\label{eq:2.44}
 \int_{-\infty}^\infty \tilde{C} (x,x',t,t' )  \bm{w}(x',t' )  \dx x' =\sigma^2 \bm{w}(x,t) \,.
\end{equation}

Since the target of the method is to retrieve advecting wavepacket structures, spatial wavenumber $k$ and temporal frequency $\omega$ can be considered to be linked through the phase velocity $c=\omega/k$. The analytic signal of the time-space correlation tensor then is modified as
\begin{equation}
        \tilde{C} (\xi,\tau)=\int_{-\infty}^\infty (1+\text{sign}(k))  S(k,\omega)  e^{i k (\xi-c \tau)}   \dx k\,.
\end{equation}
Substituting $\tilde{C}$ in Eq. \eref{eq:2.44} yields
\begin{equation}\label{eq:2.45}
\int_{-\infty}^\infty \left[\int_{-\infty}^\infty(1+\text{sign}(k))  S(k,\omega)  e^{i k (\xi-c \tau) }  \dx k \right]  \bm{w}(x',t' )  \dx x'=\sigma^2 \bm{w}(x,t) \,.
\end{equation}
After some manipulation, the equation can be rewritten as
\begin{equation}\label{eq:2.45}
\hspace{-1mm}\int_{-\infty}^\infty(1+\text{sign}(k))  S(k,\omega)  e^{i k (x-c t) }  \left[ \int_{-\infty}^\infty \bm{w}(x',t' ) e^{-i k (x'-c t' ) }   \dx x' \right]  \dx  k =\sigma^2 \bm{w}(x,t)\,.
\end{equation}

To understand the term in the square brackets of Eq. \eref{eq:2.45}, $\bm{w}$ may be considered as a wavepacket, i.e. to be a solution of the wave equation
        \begin{equation}
        \frac{\partial^2}{\partial t^2} \bm{w} - c^2 \frac{\partial^2}{\partial x^2} \bm{w}=0.
    \end{equation}
    The solution $\bm{w}$ at a generic time $t'$ is linked to the solution $\bm{w_0}$ at time $t'=0$ according to
    \begin{equation}
         \bm{w}(x',t')= \int_{-\infty}^\infty \mathcal{F}[\bm{w}_0](k') e^{i k' (x'-c t') } \dx k'\,,
    \end{equation}
    where
    \begin{equation}
        \mathcal{F}[\bm{w}_0](k')=\int_{-\infty}^\infty \bm{w}(x',0 ) e^{-i k' x' }   \dx x'\,.
    \end{equation}
    This allows to rewrite the term in square brackets of Eq. \eref{eq:2.45} as 
    \begin{equation}
        \begin{split}
                &\int_{-\infty}^\infty \bm{w}(x',t' ) e^{-i k (x'-c t' ) }   \dx x' \\
                &=\int_{-\infty}^\infty \left[\int_{-\infty}^\infty \mathcal{F}[\bm{w}_0](k') e^{i k' (x'-c t') } \dx k'\right] e^{-i k (x'-c t' ) }   \dx x' \,,\\
                &=\int_{-\infty}^\infty \mathcal{F}[\bm{w}_0](k') e^{-i (k'-k) c t' } \left[\int_{-\infty}^\infty   e^{i (k'-k) x'}   \dx x' \right]   \dx k'\,,\\
                &=\int_{-\infty}^\infty \mathcal{F}[\bm{w}_0](k') e^{-i (k'-k) c t' } \delta(k'-k)   \dx k'\,,\\
                &= \mathcal{F}[\bm{w}_0](k)= \mathcal{F}[\bm{w}](k,\omega=ck) \,,
    \end{split}
    \end{equation}
    where the last equality serves as a reminder that the generic time solution at each wavenumber $k$ evolves according to a frequency $\omega$ determined by $c$. This result, substituted in the Eq. \eref{eq:2.45} , gives

\begin{equation}
\int_{-\infty}^\infty   S(k,\omega )  e^{i k (x-c t) }   \mathcal{F}[\tilde{\bm{w}}](k,\omega) \dx k =\sigma^2 \bm{w}(x,t)\,.
\end{equation}

The eigenfunction and the Fourier transform of its analytic signal can be assumed to be
\begin{equation}
\begin{split}
\bm{w}(x,t)&=\bm{q}(k',\omega') e^{i k' (x-c' t) } \,,\\
\mathcal{F}[\tilde{\bm{w}}](k,\omega)& =\bm{q}( k',\omega' )(1+\text{sign}(k' ))\delta(k-k')\delta(c  k-\omega')\,.
\end{split}
\end{equation}

Substituting it into the eigenproblem leads to
\begin{equation}
\!\begin{multlined}\int_{-\infty}^\infty  S(k,\omega)  e^{i k (x-c t) } \bm{q}(k',\omega' ) (1+\text{sign}(k')) \delta (k-k') \delta (c k-\omega') \dx k\\ \hspace{15mm}    =\sigma^2 \bm{q}(k,\omega ) e^{ik( x-c t)}\,,  \end{multlined}
\end{equation}
which finally results in the SPOD eigenproblem for positive $k$ and $\omega/k=c$, i.e.
\begin{equation}\label{eq:SPOD}
 S(k,\omega)  \bm{q}(k,\omega) e^{i k (x-c t) }=\frac{\sigma^2}{2} \bm{q}(k,\omega) e^{i k (x-c t) }\,.
\end{equation}

A similar result would be obtained if in Eq. \eref{eq:crossspectral} the temporal direction had been taken. Eq. \eref{eq:SPOD} shows that a wavepacket structure can be recognized by properly applying translational symmetry to the problem through the Hilbert transform either in space or time, at least while a phase velocity relation is in place. This implies that, while translational symmetry can be applied, both the conventional and space-only HPOD approaches are equivalent and would be able to retrieve wavepackets. 
Notice that the presence of the integral operator in the dimension in which the Hilbert/Fourier transforms are computed, requires that such a dimension is resolved with high enough resolution. In the case of space-only \ac{HPOD}, this means that the method works while the spatial resolution is sufficiently fine to capture the displacement of the wave with enough precision according to the phase velocity. A similar condition would be applied in the temporal resolution when the conventional \ac{HPOD} problem were solved.

Also, notice that the \ac{SPOD} problem in Eq. \eref{eq:SPOD} has been derived assuming both stationary and homogeneous flow, and assuming a relation between temporal frequency and spatial wavenumber of a wavepacket through the phase velocity. In this sense, the \ac{SPOD} analysis is limited by the necessity of having a homogeneous/stationary direction in which Fourier transform can be applied. While stationarity in time is oftentimes a reasonable assumption, homogeneity along a given direction is much less common, limiting the applicability of SPOD to time-resolved sequences. In cases where homogeneity (or stationarity) cannot be assumed, a space/wavenumber (or time/frequency) method should be applied in order to instate the correct symmetries. In this sense, \ac{HPOD} provides a more suitable approach, which -- while not guaranteeing spectrally pure modes -- allows the inclusion of modulations on top of a simpler sinusoidal waveform.
The demonstration of the equivalence between the \ac{HPOD} approaches can be easily extended to cases in which the homogeneous/stationary flow assumptions drop and the HPODs do not converge to the SPOD problem any more. It is worth to highlight that the equivalence of the two \ac{HPOD} approaches only holds for advecting features, i.e. when a phase velocity relation is in place between temporal frequency and spatial wavenumber. For phenomena of different nature, the two approaches should not be expected to converge to the same solution. This, however, should not be considered a detrimental aspect of the space-only approach versus the conventional approach, but rather a feature that can be exploited, for example to target more effectively advecting features and rule out other time-dependent phenomena, thus justifying the use of the space-only \ac{HPOD} also in scenarios where temporal resolution is available.

\FloatBarrier

\section{Test cases} \label{sec:testcases}
This section proposes 3 different test cases to demonstrate the properties of the \ac{HPOD} and highlight its capabilities. The datasets are proposed in order of their complexity in terms of data-driven modal analysis. 

The first dataset is the velocity field of the laminar vortex shedding behind a cylinder extracted from 2D-\ac{DNS}. This dataset is the least complex one, in which few flow features with a clear frequency signature are present and in which temporal resolution is available. This flow has been plenty analysed by means of data-driven techniques --\ac{POD} being a primary example--, making it a perfect didactic example with a well-known and easily-recognizable flow model. As such, it allows demonstrating the most basic properties of the \ac{HPOD}, in particular its capability to deliver spatiotemporally coherent structures, already organized in complex-valued modes, which do not require \emph{a posteriori} pairing as the \ac{POD} ones. Moreover, its robustness to the lack of temporal resolution in the space-only implementation can be introduced. 

The second dataset is the velocity field of a turbulent jet flow extracted from a \ac{LES}. This dataset represent a more complex scenario than the previous one, since -- despite the presence of advective flow features -- a large wealth of turbulent scales is present and there is no clear frequency signature. This dataset is used to demonstrate the capabilities of the \ac{HPOD} to extract advective structures also in a very challenging scenario characterized by a broadband spectral content, whereas other techniques can be subjected to mode mixing. Additionally, it allows to emphasize the analytic signal nature of the HPOD modes and its effect of their frequency/wavenumber content.

Finally, the third dataset is a collection of flow field snapshots of a turbulent jet measured with planar \ac{PIV}. This dataset provides similar challenges with respect to the previous case in terms of turbulence scales and broadband spectral content, adding on top of it measurement uncertainty and temporally-unresolved data. This last dataset is used to demonstrate the usefulness of the space-only \ac{HPOD} in retrieving spatial modes interpretable as advective flow structures (similar to the ones obtained in the \ac{LES} case) even using data from experiments with no temporal information available.

\subsection{DNS of a laminar vortex street in the wake of a cylinder}\label{sec:cyltestcase}

The conventional and space-only \ac{HPOD} are tested on a numerical database obtained from the 2D-\ac{DNS} of the flow around a cylinder of diameter $D$ at $Re=100$. The flow around the cylinder has been computed through the OpenFOAM solver pimpleFoam. The dataset used in this test case includes $10\,800$ time-resolved snapshots spanning a cropped domain of $10.6D \times 4.8D$ corresponding to $91 \times 201$ mesh points. The temporal separation between frames corresponds to a advective time $\Delta t U_\infty/D=0.0139$, resulting in a time-resolved sequence. This database has been selected, since the laminar vortex street in the wake of the cylinder offers a clear periodic signature, which can help clarify the potential of the space-only application of the \ac{HPOD}. Additionally, since the original sequence is time-resolved, the conventional \ac{HPOD} (applied in time) can be implemented, providing a point of comparison between the 2 distinct implementations of \ac{HPOD}. 

\begin{figure*}
\centering
\includegraphics[width=\linewidth]{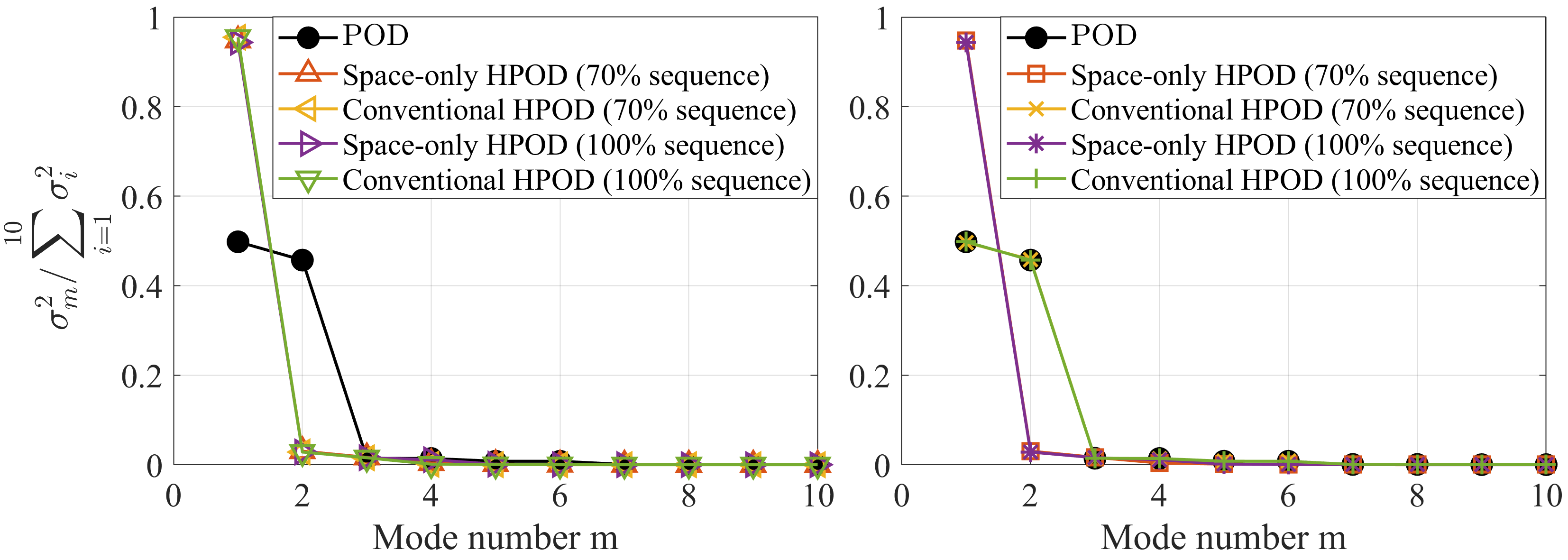}
\caption{ Energy content of each mode versus the total energy content of the sequence for the \ac{HPOD} implementations and \ac{POD}: (left) original time-resolved sequence; (right) sequence shuffled in time.\label{fig:cyl_spect}}
\end{figure*}

Fig. \ref{fig:cyl_spect}(left) reports the ratio of the first 10 squared singular values, representing the energy content in each \emph{structure}, versus the total energy content of the sequence for the \ac{HPOD} implementations in time and in space (along the $x$ direction) for the original time-resolved sequence. To exemplify the implication of the   edge effects described in Section \ref{sec:complexification}, the \ac{HPOD} has been applied both considering a 15\% data removal at the end of each side of the signal (either along the $x$ direction or $t$ depending on the implementation), i.e. retaining information of 70\% of the sequence, as well as retaining the 100\% of the sequence. The traditional \ac{POD} has been included for comparison. The distribution of the energy shows that, in this case, both the \ac{HPOD} implementations retrieve a more compact decomposition than \ac{POD}, predicting only a single structure containing most of the energy, while the \ac{POD} -- as expected -- splits the energy in the first 2 structures. Discrepancies between the two implementations can be observed only for low-energy structures and they are typically negligible. The distribution of energy seems to be even less altered when part of the sequence is filtered out.
Fig. \ref{fig:cyl_spect}(right) reports the same energy content distribution computed for the time-resolved sequence being shuffled to mimic non-time-resolved data. As expected, while the space-only \ac{HPOD} retain the same energy distribution as in the unshuffled case, the conventional implementation in time cannot correctly complexify the sequence, converging, therefore, towards the standard \ac{POD}. This phenomenon has already been reported by \cite{kriegseis2021hilbert}.

\begin{figure*}
\centering
\includegraphics[width=0.95\linewidth, trim= {0 0.7cm 0 0}]{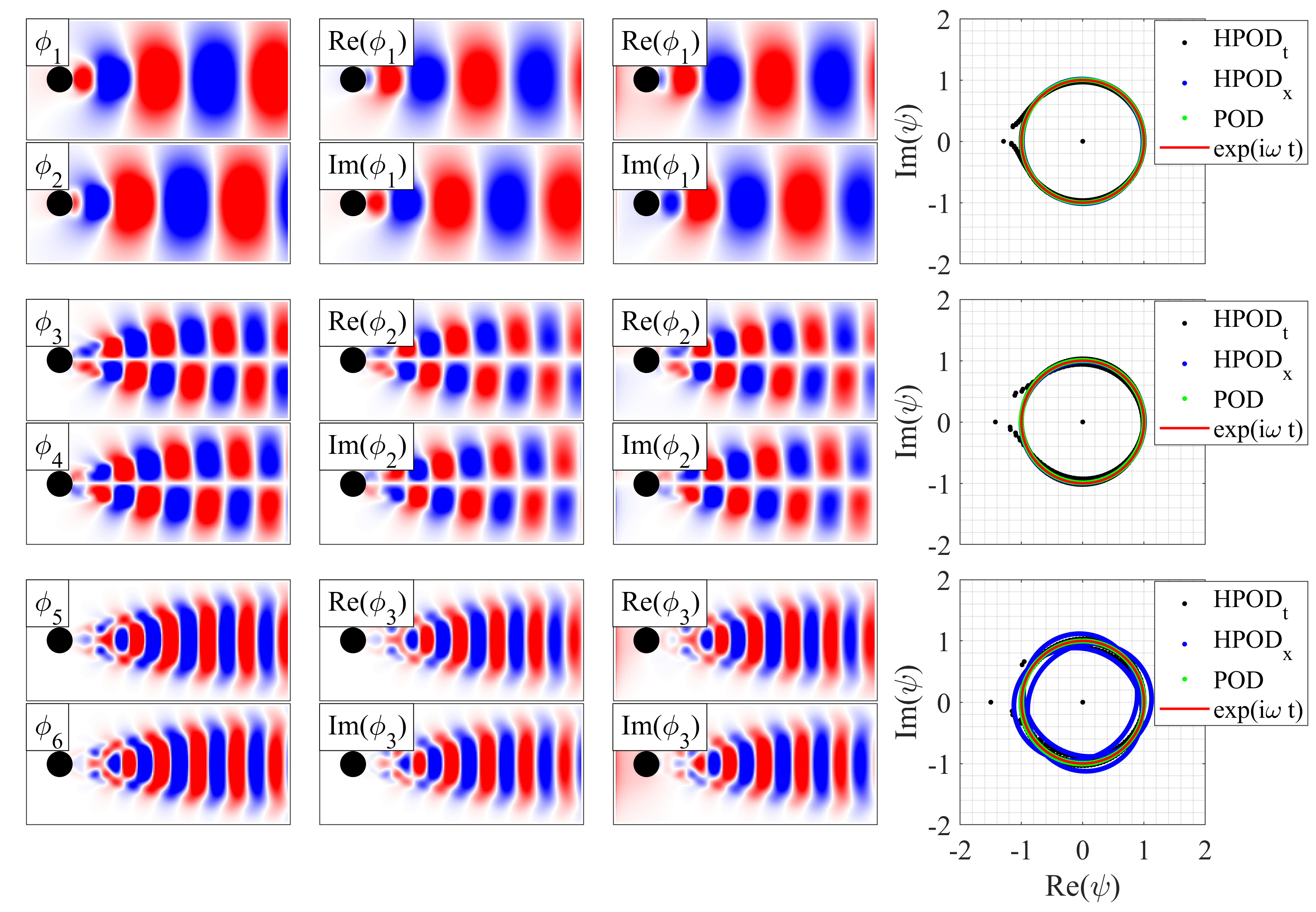}
\caption{Comparison between \ac{POD} and \ac{HPOD} modes -- computed on the original time-resolved sequence, 100\% of the sequence: first 3 pairs of spatial \ac{POD} modes (1\textsuperscript{st} column); first 3 spatial modes of the conventional \ac{HPOD} (2\textsuperscript{nd} column); first 3 spatial modes of the space-only \ac{HPOD}  (3\textsuperscript{rd} column); phase plot of the temporal modes for \ac{HPOD} implementations and for \ac{POD} pairs (4\textsuperscript{th} column).}\label{fig:cyl_mod_or_100}

\centering
\includegraphics[width=0.95\linewidth, trim= {0 0.7cm 0 0}]{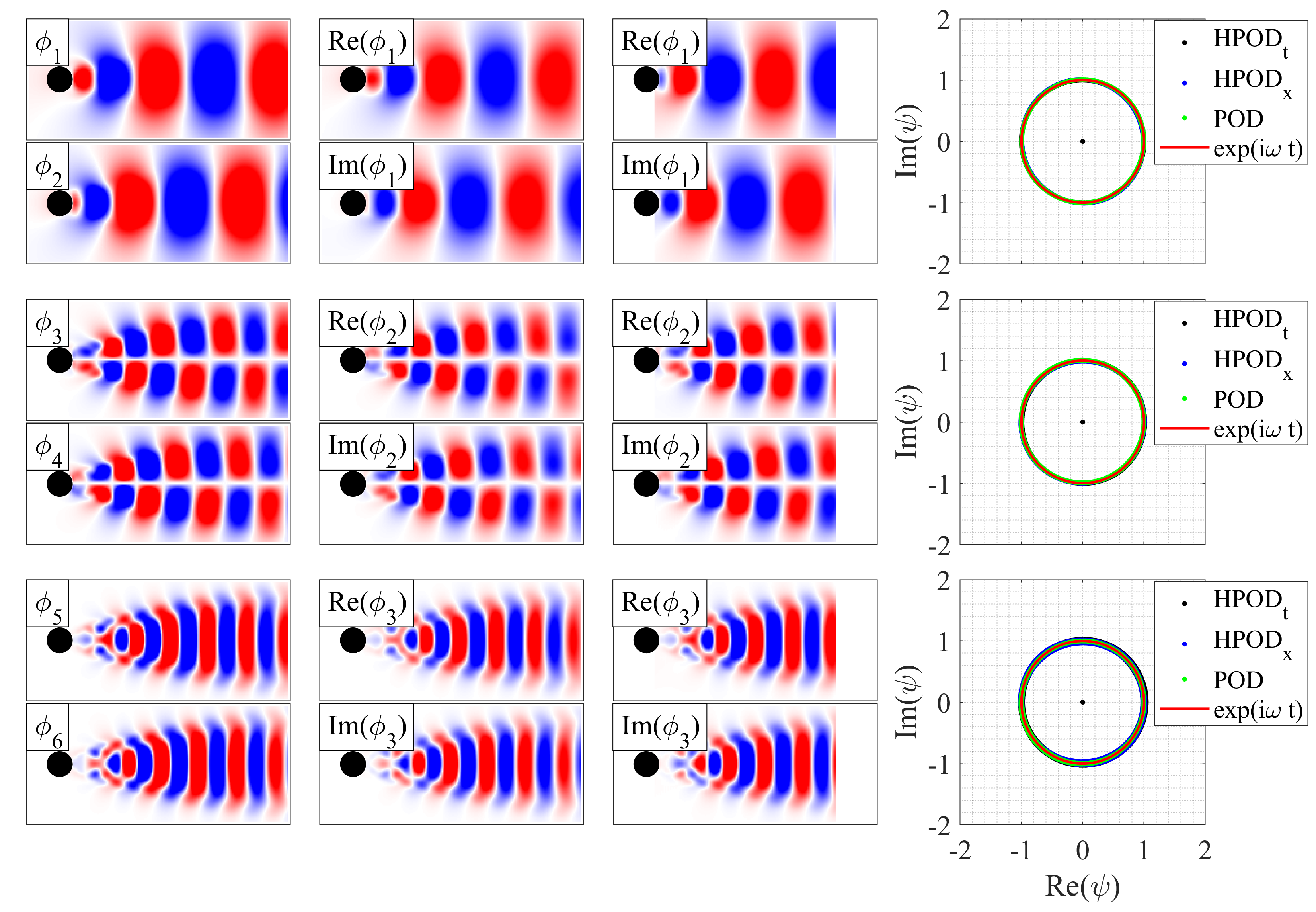}
\caption{Comparison between \ac{POD} and \ac{HPOD} modes -- computed on the original time-resolved sequence, 70\% of the sequence: first 3 pairs of spatial \ac{POD} modes (1\textsuperscript{st} column); first 3 spatial modes of the conventional \ac{HPOD} (2\textsuperscript{nd} column); first 3 spatial modes of the space-only \ac{HPOD} (3\textsuperscript{rd} column); phase plot of the temporal modes for \ac{HPOD} implementations and for \ac{POD} pairs (4\textsuperscript{th} column).}\label{fig:cyl_mod_or_70}
\end{figure*}

While the distribution of energy and the compactness of the decomposition are relevant features, the main interest in this test case is to show the capabilities of HPOD to retrieve advective flow structures. In this sense, the wake behind of a cylinder is a perfect test case for didactic purposes: the vortex street is represented by travelling waves of vorticity propagating in the flow direction with little to no temporal modulation. For simple flow scenarios not affected by mode mixing, the standard POD, as most of the readers had already experienced, identify for each wave a pairs of modes sharing a quadrature phase relation both in space and time. These modes can be paired \emph{a posteriori} as real and imaginary parts of an oscillatory model, which represent the benchmark against which the HPOD should be compared.
Figs. \ref{fig:cyl_mod_or_100} and \ref{fig:cyl_mod_or_70} report the first 3 \emph{structures} retrieved by the conventional implementation of the \ac{HPOD} in time (2\textsuperscript{nd} column, referred to as HPOD\textsubscript{t} in here) and for the space-only implementation of the \ac{HPOD} in the advection direction $x$ (3\textsuperscript{rd} column, referred to as HPOD\textsubscript{x} in here) for the time-resolved original sequence respectively without data removal (100\% of the sequence,  Fig. \ref{fig:cyl_mod_or_100}) and with data removal (70\% of the sequence, Fig. \ref{fig:cyl_mod_or_70}). The complex-valued spatial modes are reported in terms of their real and imaginary parts. The first 6 spatial modes of the POD, grouped in pairs, are reported for comparison in the 1\textsuperscript{st} column. The 4\textsuperscript{th} column reports the temporal mode of the \ac{HPOD} implementations and of the POD mode pairs in the form of a phase plot (real vs. imaginary part for the \ac{HPOD} and cyclogram between paired modes for the \ac{POD}). 
The spatial modes retrieved by the 2 \ac{HPOD} implementations look extremely similar in amplitude and wavelength in the $x$ direction, differing only for being in phase opposition, as shown by the discrepancy in their imaginary part. This discrepancy is by construction and does not affect the physical behaviour of the retrieved structures, as will be fully clarified for the next testcase. The retrieved complex-valued spatial modes corresponds to mode pairs of the \ac{POD}, which clearly represent pairs of modes in phase quadrature in space. In terms of phase plots of temporal modes, the \ac{POD}, as expected for a von Kármán street, retrieves mode pairs, which behave temporally as pure sinusoid with a phase shift of $\pi/2$: each temporal mode pair, represented as the real and imaginary parts of a single complex-valued mode, form a Lissajous figure corresponding to the unitary circle $\text{exp}(i\omega t)$, represented in red in the plots. The complex-valued temporal modes from the \ac{HPOD}s tend to follow the same unitary circle, which is compatible with spectrally-pure modes in time.

An important point needs to be highlighted when looking at the differences between full sequence and reduced sequence results: for the full sequence, distortions of the modes appear. These distortions are typically relevant in the boundary of the direction in which the Hilbert transform has been implemented (the time in the conventional implementation and the direction $x$ in the space-only implementation), which are compatible with the edge effects of the transform. For low-energy modes, this distortion seems to propagate also into other directions, as shown for the space-only \ac{HPOD} in Fig. \ref{fig:cyl_mod_or_100} row 3, column 4, where the distortion propagates also in time indicating that the mode is corrupted. These distortion effects seem to be completely suppressed if 15\% of the signal is removed from the signal end after the Hilbert transform. Obviously, this signal reduction is reflected in the reduction of the domain in which the mode is retrieved, which in the space-only implementation results in a shrinkage of the spatial domain in the $x$ direction. It must be remarked that this 15\% signal reduction has been selected on an empirical basis as a conservative measure to ensure edge effects are removed completely. An accurate treatment of Hilbert transform edge effect is outside the scope of the present work. Less conservative criteria, or even other edge-effect treatment techniques, can be implemented to reduce shrinkage of the signal.

\begin{figure*}
\centering
\includegraphics[width=0.95\linewidth, trim= {0 0.7cm 0 0}]{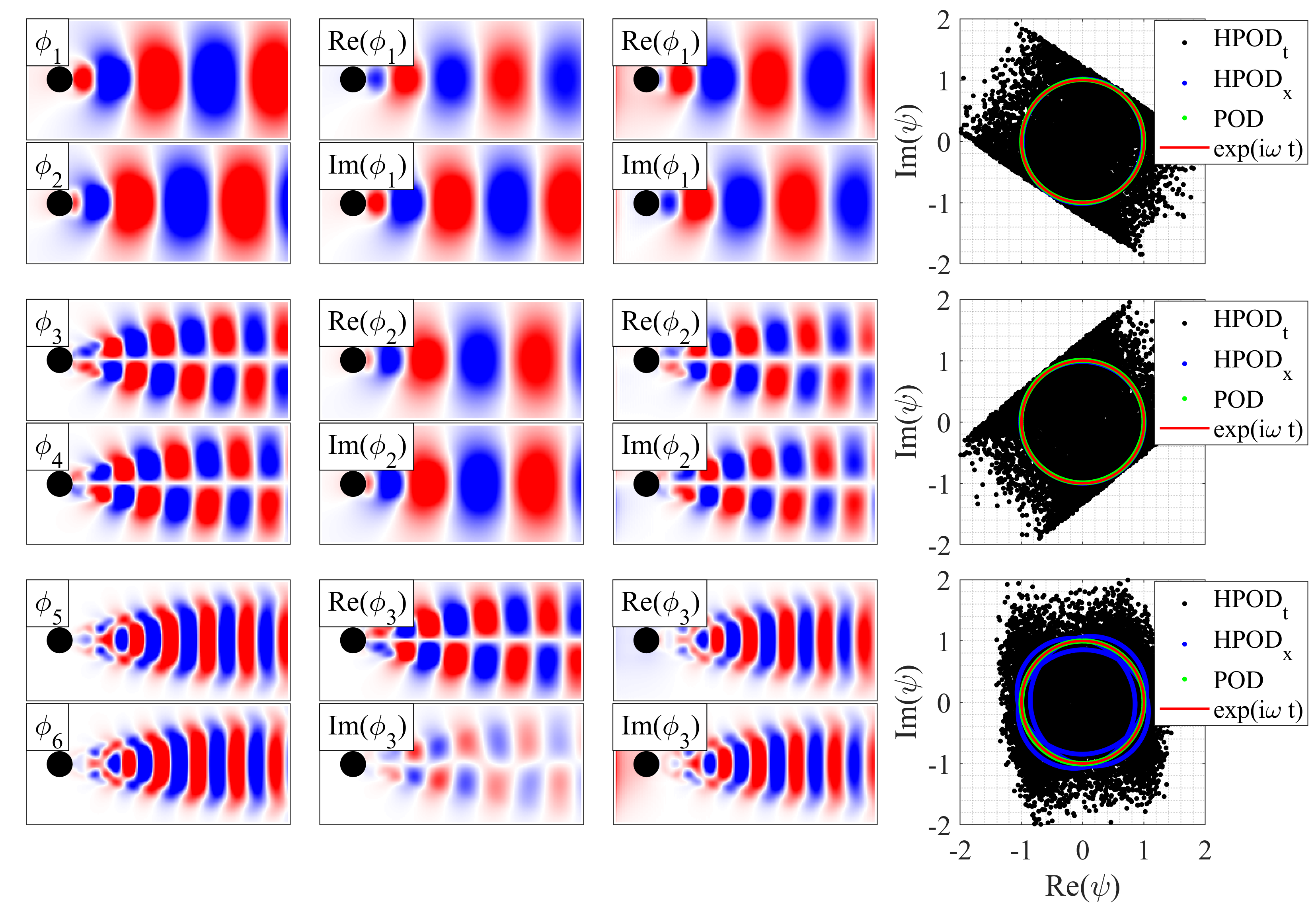}
\caption{Comparison between \ac{POD} and \ac{HPOD} modes -- computed on the shuffled sequence, 100\% of the sequence: first 3 pairs of spatial \ac{POD} modes (1\textsuperscript{st} column); first 3 spatial modes of the conventional \ac{HPOD} (2\textsuperscript{nd} column); first 3 spatial modes of the space-only \ac{HPOD} (3\textsuperscript{rd} column); phase plot of the temporal modes for \ac{HPOD} implementations and for \ac{POD} pairs (4\textsuperscript{th} column).}\label{fig:cyl_mod_sh_100}

\centering
\includegraphics[width=0.95\linewidth, trim= {0 0.7cm 0 0}]{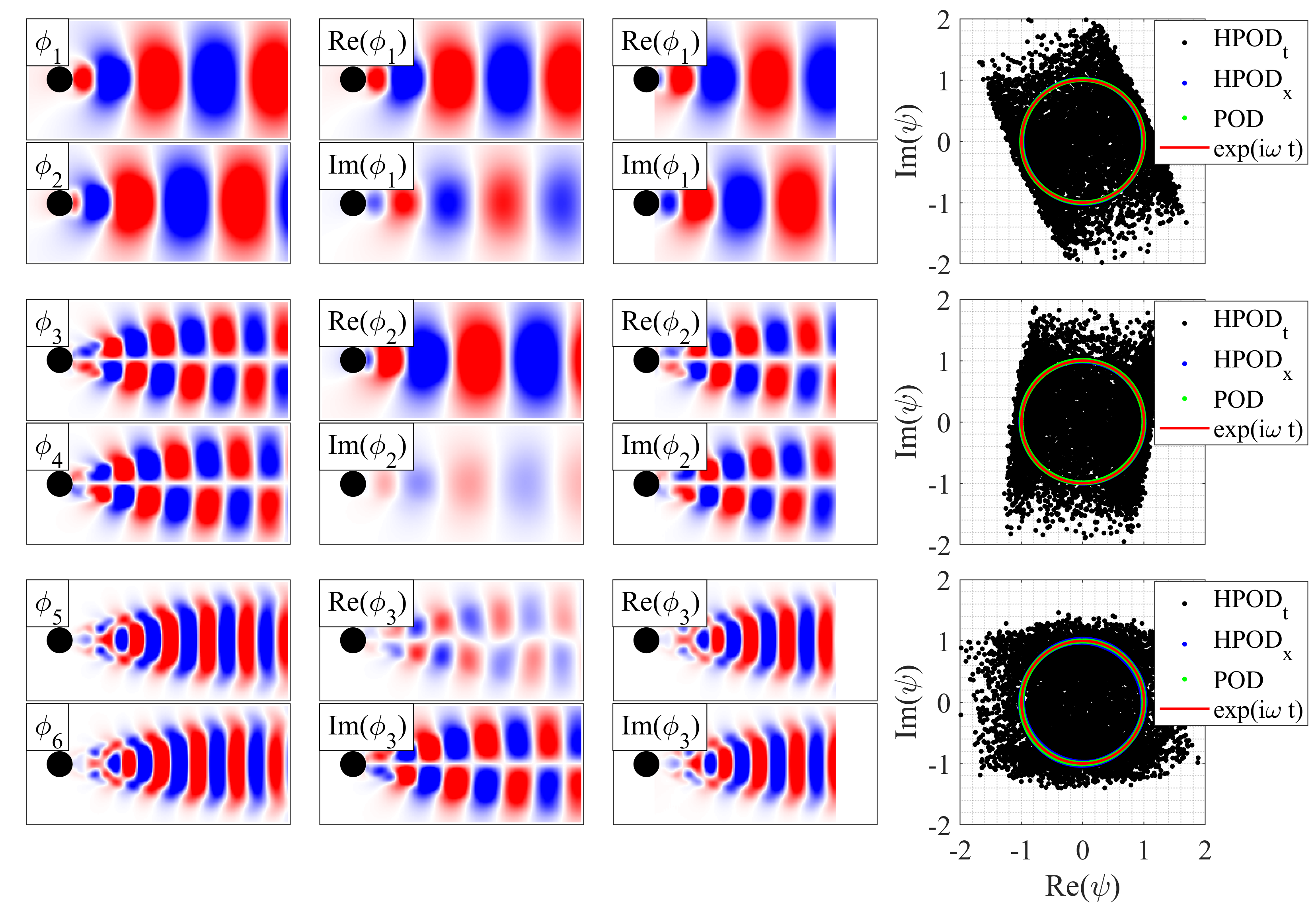}
\caption{Comparison between \ac{POD} and \ac{HPOD} modes -- computed on the shuffled sequence, 70\% of the sequence: first 3 pairs of spatial \ac{POD} modes (1\textsuperscript{st} column); first 3 spatial modes of the conventional \ac{HPOD} (2\textsuperscript{nd} column); first 3 spatial modes of the space-only \ac{HPOD} (3\textsuperscript{rd} column); phase plot of the temporal modes for \ac{HPOD} implementations and for \ac{POD} pairs (4\textsuperscript{th} column).}\label{fig:cyl_mod_sh_70}
\end{figure*}

The \ac{HPOD} is repeated for the temporally shuffled sequence of velocity fields in order to simulate a non-time-resolved dataset. Figs. \ref{fig:cyl_mod_sh_100} and \ref{fig:cyl_mod_sh_70} shows the first 3 \emph{structures} retrieved by the conventional and space-only implementations of  \ac{HPOD} respectively, without data removal (100\% of the sequence, Fig. \ref{fig:cyl_mod_sh_100}) and with data removal (70\% of the sequence, Fig. \ref{fig:cyl_mod_sh_70}). As expected, the space-only \ac{HPOD} is unaffected by the sequence not being time-resolved and still provides results that match the \ac{POD} pairs both in terms of spatial and temporal modes. The conventional \ac{HPOD}, instead, cannot properly complexify the dataset due to the lack of resolution in the temporal direction. The modes it retrieves are, therefore, converging towards \ac{POD} modes both in terms of space and time. This is evident by the fact that each complex-valued temporal mode does not show a clear quadrature phase relation between its real and imaginary parts, as in the previous case. Also, spatial modes with the same spatial frequency content appear, i.e. mode pairs as in the \ac{POD} approach, showing that \ac{HPOD} is not capable of recognizing them as part of a single complex-valued mode. It is worth highlighting that despite the modes tends to converge towards the \ac{POD} ones, they retain their complex-valued nature, having a non-null imaginary part. 

\subsection{LES of a turbulent jet}

The performance of the conventional and space-only \ac{HPOD} are further tested in a more complex scenario, characterized by a broadband spectral content but still dominated by advecting flow structures. A velocity field database has been extracted from the high-fidelity \ac{LES} of a turbulent jet at $M=U_j/a=0.9$ and $Re_j=U_j D/\nu\approx10^6$ provided by \cite{towne2023database}. This flow is characterized by travelling wavepacket forming in the shear layer -- as shown through SPOD analysis by \cite{towne2023database} -- which are not shed at a fixed frequency but rather posses a broadband spectrum. This dataset allows to demonstrate the capability of HPOD to retrieve spatiotemporally coherent features characterized by modulation (and intermittency) both in space and time, and thus profoundly differing from spectrally-pure SPOD modes. The database is constituted by $10\,000$ time-resolved snapshots sampled at $\Delta t =0.2 D/a$, where $a$ is the speed of sound and $D$ is the nozzle diameter. The data are available over a structured cylindrical grid spanning 30 D in the axial direction and 6 D in the radial one, containing (656,138,128) points in the axial, radial and azimuthal directions. For the application of the \ac{HPOD}, a single plane at a fixed azimuthal coordinate has been extracted for each snapshot. 

\begin{figure*}
\centering
\includegraphics[width=\textwidth]{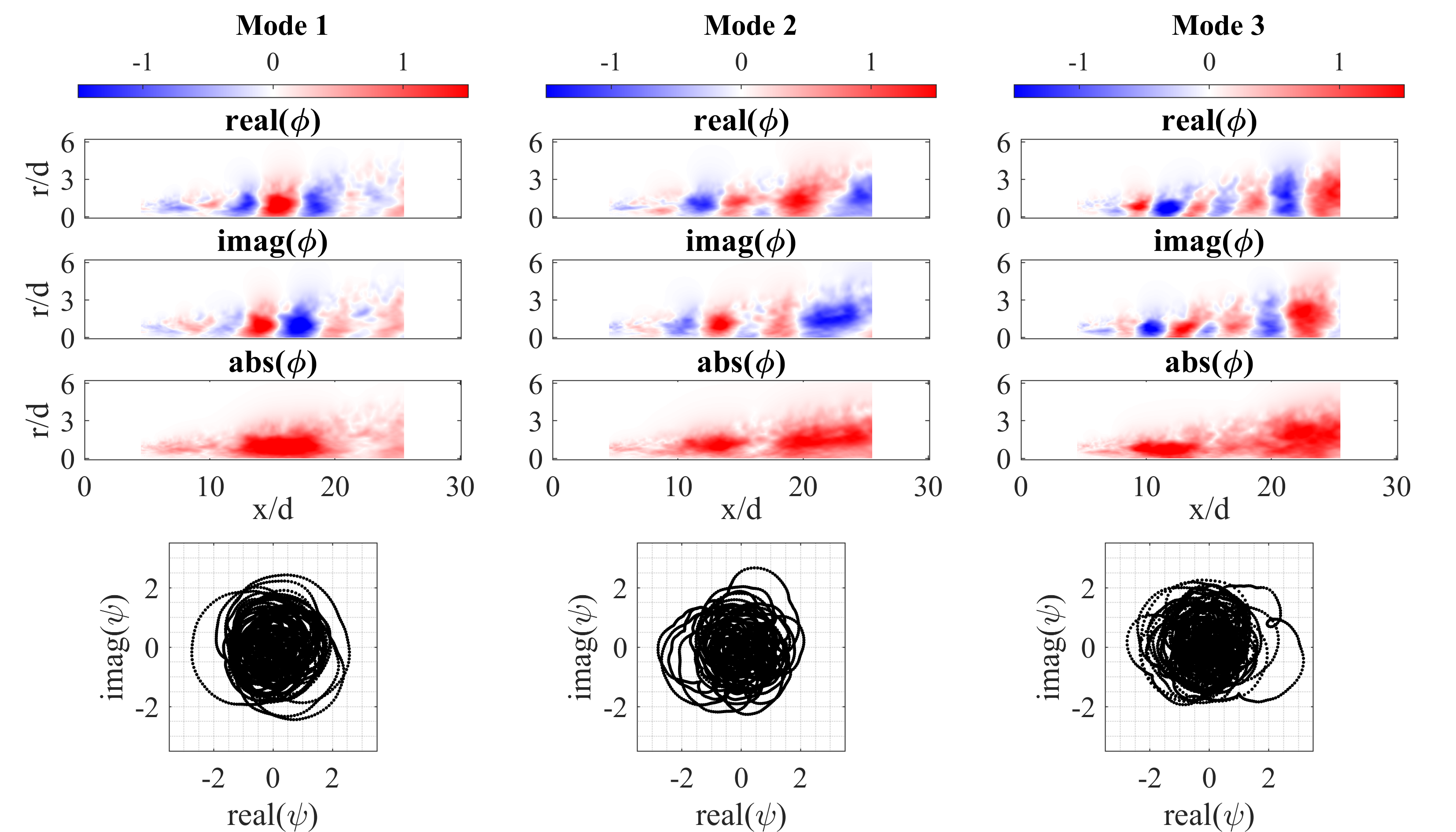}
\caption{Radial velocity component of the complex-valued spatial modes of the \ac{HPOD} performed in the temporal direction: (left) 1\textsuperscript{st} mode, (centre) 2\textsuperscript{nd} mode, (right) 3\textsuperscript{rd} mode. From top to bottom: real part of the spatial mode; imaginary part of the spatial mode; absolute value of the spatial mode; phase plot of the real versus imaginary part of the temporal mode. }\label{fig:tHPOD_Jet_LES}

\centering
\includegraphics[width=\textwidth]{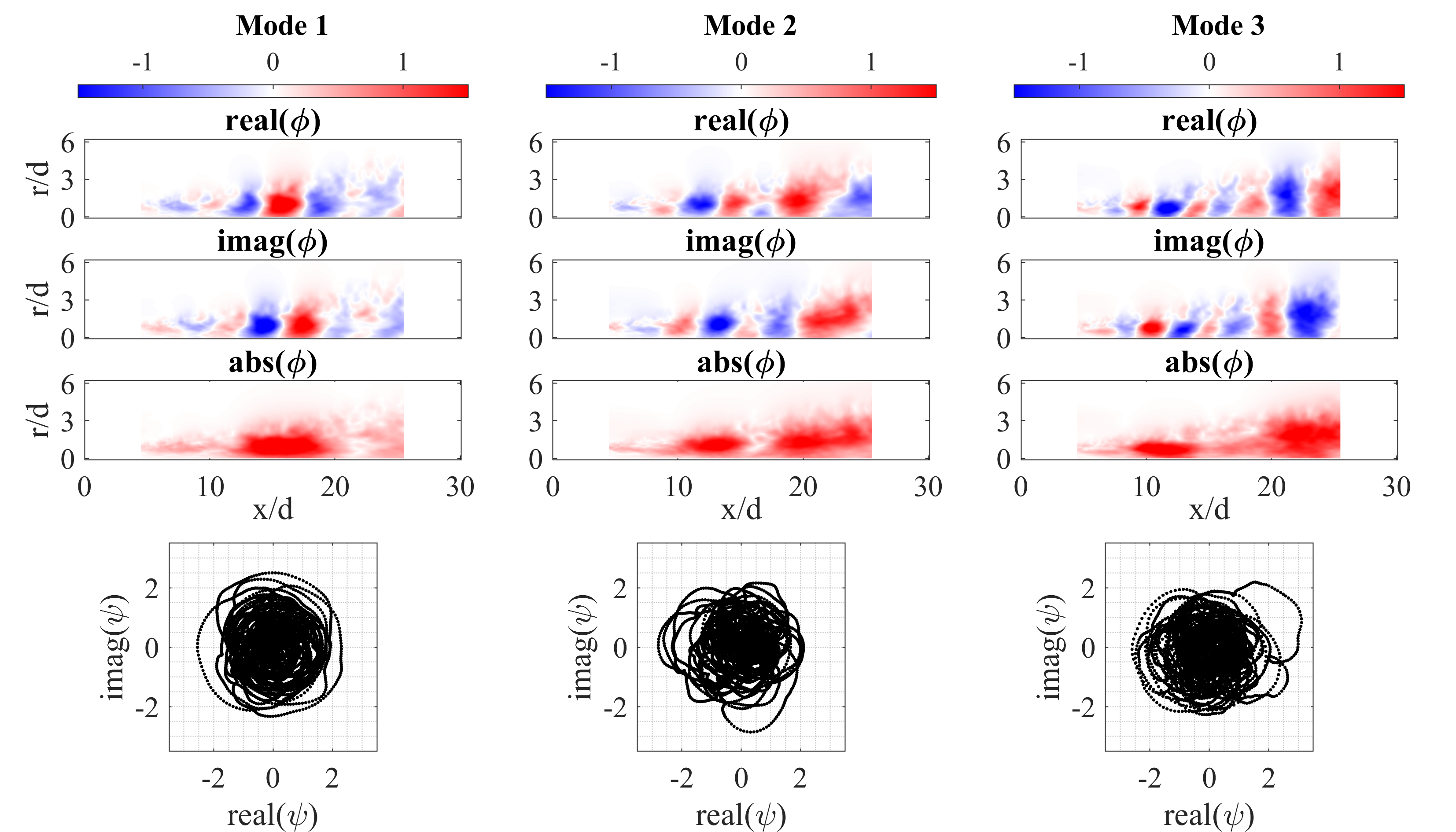}
\caption{Radial velocity component of the complex-valued spatial modes of the \ac{HPOD} performed in the advective direction: (left) 1\textsuperscript{st} mode, (centre) 2\textsuperscript{nd} mode, (right) 3\textsuperscript{rd} mode. From top to bottom: real part of the spatial mode; imaginary part of the spatial mode; absolute value of the spatial mode; phase plot of the real versus imaginary part of the temporal mode. }\label{fig:xHPOD_Jet_LES}
\end{figure*}

The conventional \ac{HPOD} applied in the temporal direction has been compared to the space-only \ac{HPOD} performed along the axial direction. In order to provide a fair comparison, 15\% of the data at each end of the signal in both the temporal and spatial directions have been removed. The first 3 complex-valued structures computed with both approaches are presented in Fig. \ref{fig:tHPOD_Jet_LES} and Fig. \ref{fig:xHPOD_Jet_LES}, respectively. These structures correspond to approximately 8.1\%, 7.2\% and 6.0\% of the turbulent kinetic energy contained in the dataset for the conventional HPOD and to approximately 8.4\%, 7.5\% and 6.2\% for the space-only HPOD. Given their complex-valued nature, the spatial modes have been represented with their real (1\textsuperscript{st} row), imaginary (2\textsuperscript{nd} row) and absolute (3\textsuperscript{rd} row) values, while temporal modes have been represented as a phase plot of the real value versus the imaginary value (4\textsuperscript{th} row). Both the \ac{HPOD} applied in time and in space show similar spatial modes for corresponding structures, i.e. a wavepacket structure with similar spatial frequency content, where the real and imaginary components are shifted by $\pi/2$ in the advection direction. The main difference between the patterns of the \ac{HPOD} performed in time and space is the phase of the modes. For example, the real part of the 1\textsuperscript{st} mode is almost identical, while the imaginary part has opposite signs, suggesting that the modes are in phase opposition. Similar behaviours can be observed also for the 2\textsuperscript{nd} and 3\textsuperscript{rd} modes, showing that the spatial modes delivered by the space-only HPOD are approximately the complex conjugate of the spatial modes delivered by the conventional HPOD (manifesting itself as a $\pi$ shift in the complex-valued mode).
The intensity of the flow field wave-like oscillations present an almost identical modulation along the streamwise direction, which can be better appreciated from the absolute value of the mode. In particular, the 1\textsuperscript{st} mode shows a distributed peak between $x/d=10$ and $x/d=20$, the 2\textsuperscript{nd} mode 2 peaks at about $x/d=13$ and $x/d=21$, while the 3\textsuperscript{rd} mode shows 2 peaks at $x/d\approx12$ and $x/d\approx23$. A similar modulation pattern is also observable on the \ac{SPOD} modes recovered by \cite{towne2018spectral} for a turbulent jet at $M=0.4$.

\begin{figure*}
\centering
\includegraphics[width=\textwidth]{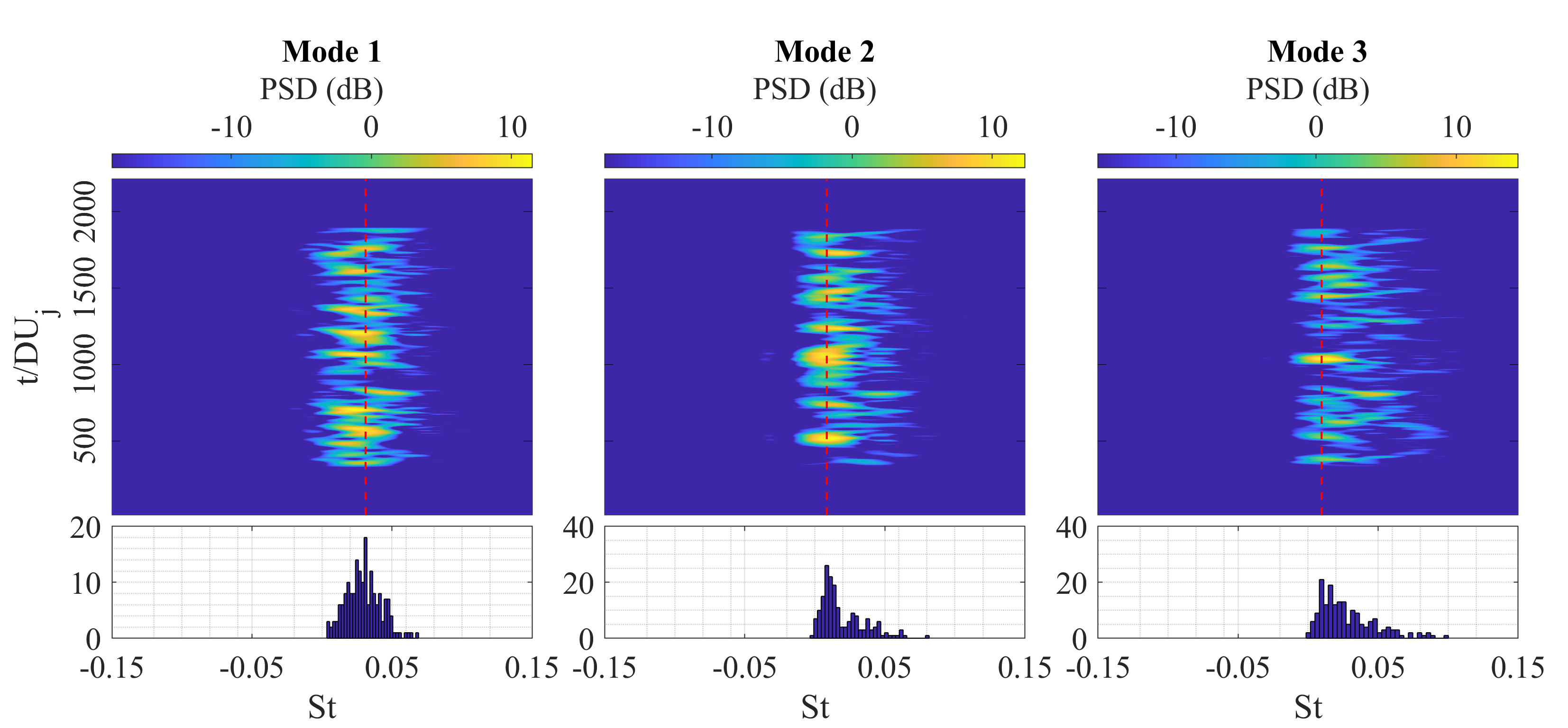}
\caption{Spectrogram of the complex-valued temporal modes of the \ac{HPOD} performed in the temporal direction (top) and histogram of the peak frequency through time (bottom). From left to right: (left) 1\textsuperscript{st} mode, (centre) 2\textsuperscript{nd} mode, (right) 3\textsuperscript{rd} mode.}\label{fig:tHPOD_Jet_LES_TF}

\centering
\includegraphics[width=\textwidth]{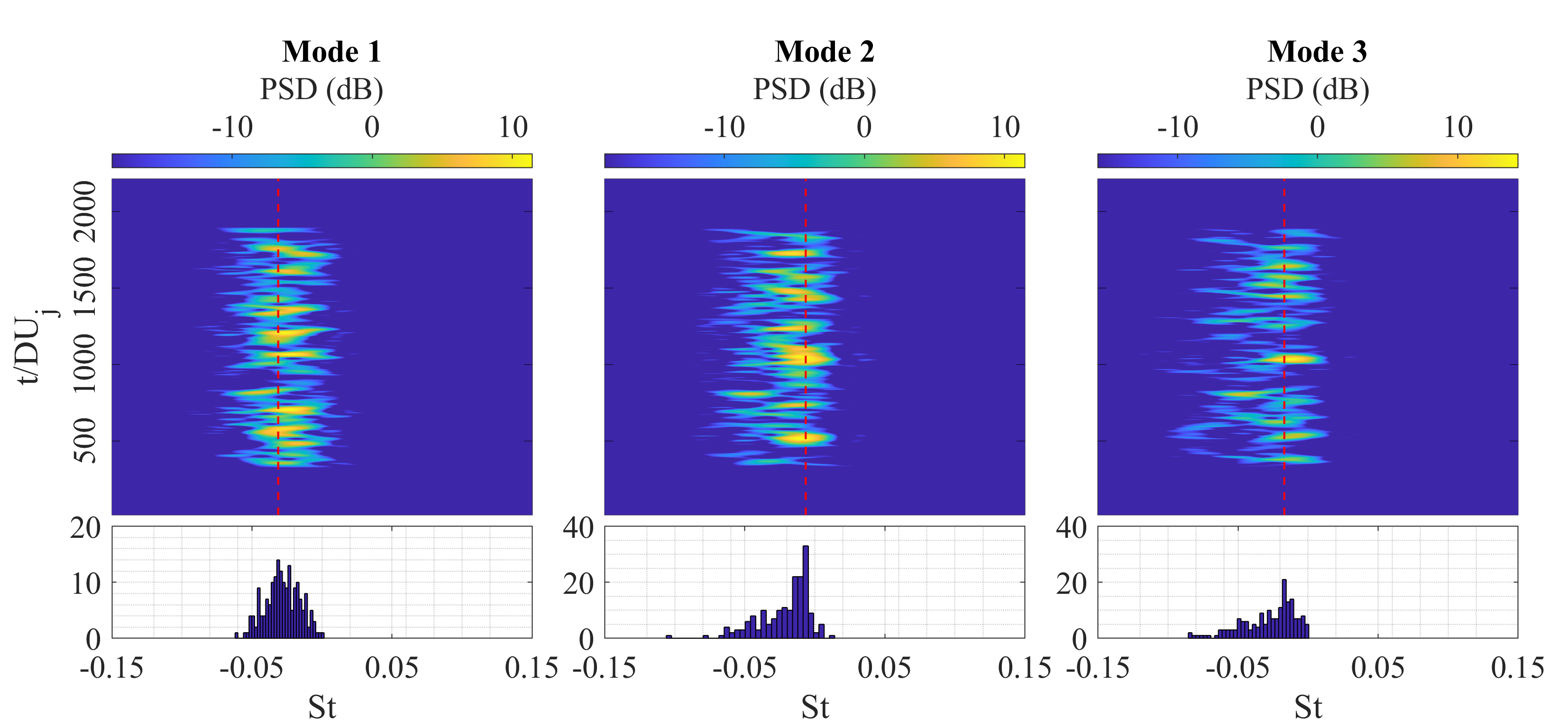}
\caption{Spectrogram of the complex-valued temporal modes of the \ac{HPOD} performed in the advective direction (top) and histogram of the peak frequency through time (bottom). From left to right: (left) 1\textsuperscript{st} mode, (centre) 2\textsuperscript{nd} mode, (right) 3\textsuperscript{rd} mode.}\label{fig:xHPOD_Jet_LES_TF}
\end{figure*}

The phase plots of the temporal modes reported in the 4\textsuperscript{th} row of Figs. \ref{fig:tHPOD_Jet_LES} and \ref{fig:xHPOD_Jet_LES} show that the retrieved jet modes do not distribute around a unitary circle $\text{exp}(i\omega t)$. This means that these modes are not spectrally pure, differently to what observed for the cylinder wake, but are instead characterized by modulation in both amplitude and frequency. To have a better insight into the behaviour of these temporal modes, a time-frequency analysis is proposed in Figs. \ref{fig:tHPOD_Jet_LES_TF} and \ref{fig:xHPOD_Jet_LES_TF}. The spectrograms have been computed through windowing the signal with rectangular windows with 75\% overlap to obtain 248 signal segments with equal span centred at different instants. Short-time Fourier transform has been computed on each signal segment to provide its frequency content at a given instant. The values of the \ac{PSD} are plotted in a scale spanning 30dB from the maximum value. Notice that, due to the complex-valued nature of the modes, the power spectrum is not even, i.e. symmetric with respect to the zero frequency. It is worth highlighting that these modes have either the negative- or the positive-frequency side of the spectrum equal to zero. This behaviour confirms that the modes are analytic in nature.

The  time-frequency analysis reveal that the spectral content of these modes varies with time but appears band-limited in frequency. The temporal modes alternate instants in which (i) the mode is active and its spectral content peaks at a given frequency, and instants in which (ii) the spectral content goes to zero, i.e. the mode is inactive. This behaviour indicates that the retrieved modes are intermittent. Apart from the modulation in amplitude, showing different energy content in time, the temporal modes are also affected by a modulation in frequency, with slight shifts of the frequency content through time. To further stress this point, the histogram of the peak frequency in each temporal window analysed in the spectrogram is reported. Despite the peak frequency changes through time, statistically it seems to distribute around a specific frequency around which the mode is modulated in frequency. 

\begin{figure*}
\centering
\includegraphics[width=\textwidth]{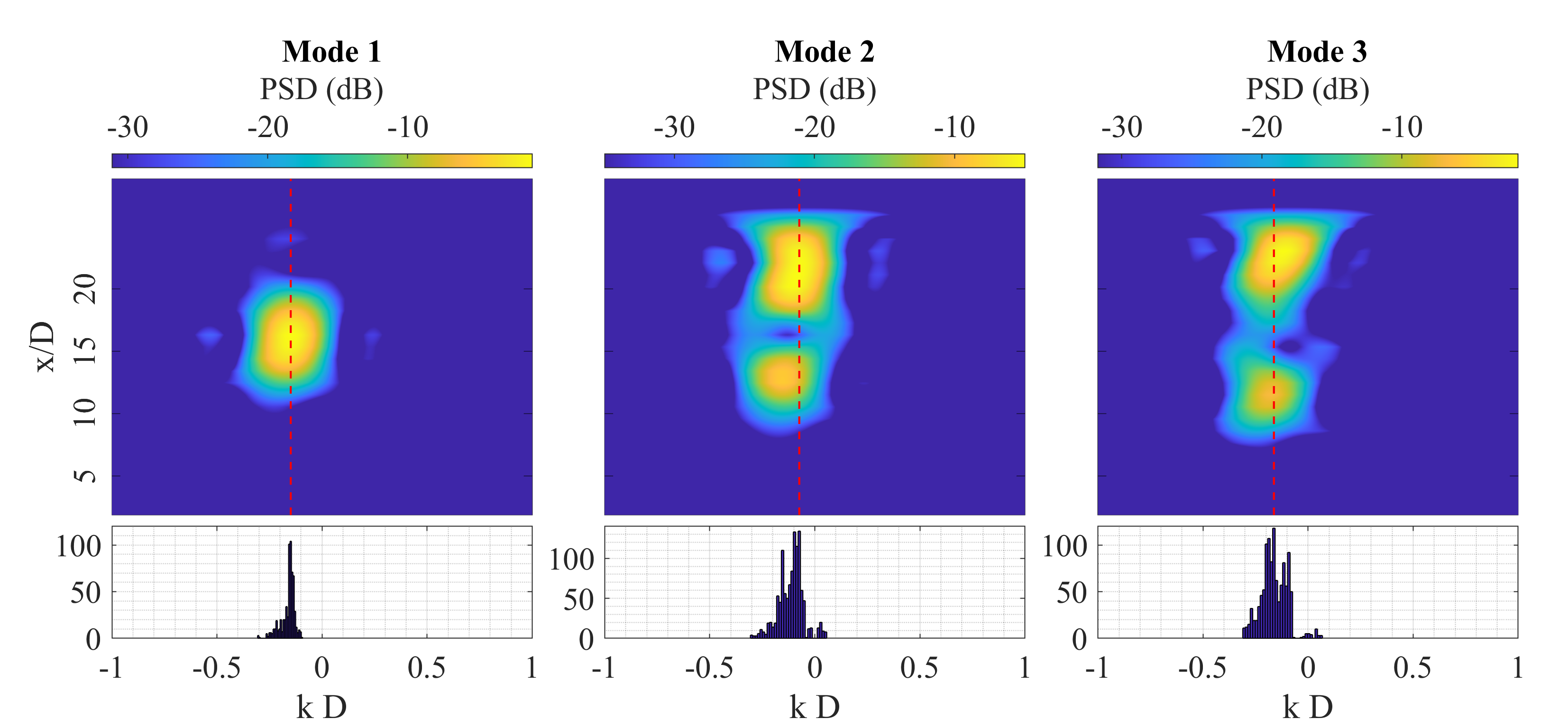}
\caption{Spectrogram along $x$ of the complex-valued spatial modes of the conventional \ac{HPOD} (top) and histogram of the peak spatial wavenumber through the advective direction (bottom). From left to right: (left) 1\textsuperscript{st} mode, (centre) 2\textsuperscript{nd} mode, (right) 3\textsuperscript{rd} mode.}\label{fig:tHPOD_Jet_LES_SW}

\centering
\includegraphics[width=\textwidth]{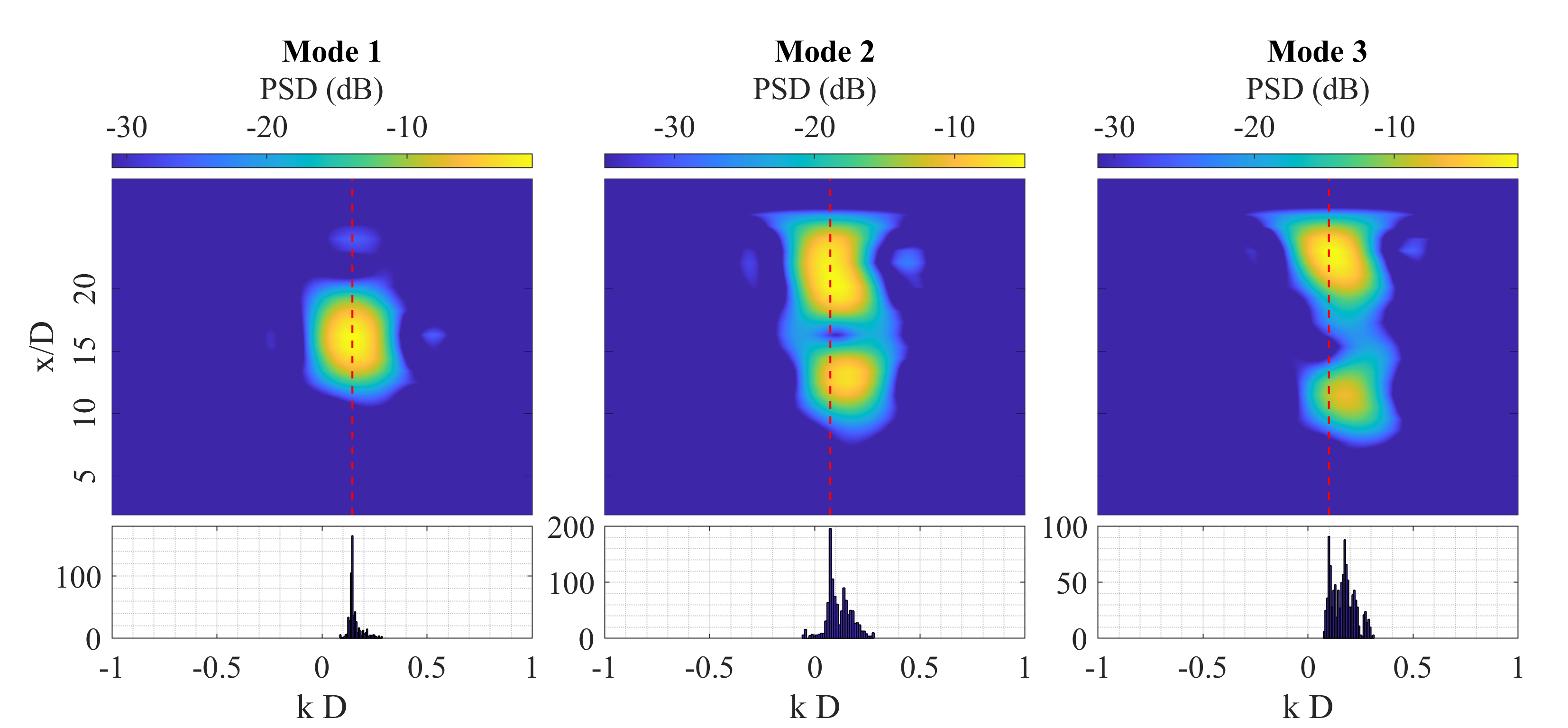}
\caption{Spectrogram along $x$ of the complex-valued temporal modes of the space-only \ac{HPOD}  (top) and histogram of the peak spatial wavenumber through the advective direction (bottom). From left to right: (left) 1\textsuperscript{st} mode, (centre) 2\textsuperscript{nd} mode, (right) 3\textsuperscript{rd} mode.}\label{fig:xHPOD_Jet_LES_SW}
\end{figure*}

A similar space-wavenumber analysis is performed on the spatial modes delivered by the conventional and space-only \ac{HPOD} in Figs. \ref{fig:tHPOD_Jet_LES_SW} and \ref{fig:xHPOD_Jet_LES_SW}. Rows along $x$ of the vector field have been extracted to obtain a signal for each radial position $r$. For each signal, the spectrograms have been computed through windowing with rectangular windows with 75\% overlap to obtain 29 signal segments with equal span centred at different $x$ positions. Short-time Fourier transform has been computed on each signal segment to provide its frequency content at a given spatial location. The values of the \ac{PSD} averaged along the radial position are plotted in a scale spanning 30dB from the maximum value. Similarly to the temporal modes, also spatial modes have either the negative- or the positive-wavenumber side of the spectrum equal to zero. This behaviour confirms that the spatial patterns as well are analytic in nature. Additionally, also the spatial wavenumber content of the spatial patterns is band-limited, even if characterized by modulation in amplitude and wavenumber depending on the axial position. This is better visualized through the histogram of the peak wavenumber in each spatial location (in both $x$ and $r$), which reveals a distribution around a specific spatial wavenumber around which the mode shows a more or less strong modulation in wavenumber. As the mode number is increased, this peak wavenumber value is shown to increase. This behaviour is to be expected, since typically large scale flow structures carry higher kinetic energy, and thus will be ranked first by the HPOD.

It is worth highlighting that the spectrograms of the temporal and spatial modes computed by both the conventional and space-only \ac{HPOD} are almost perfectly symmetric with respect to the zero frequency/wavenumber, showing almost the same spectral content apart from the sign. In particular, the temporal spectral content for the conventional \ac{HPOD} only occupies the positive-frequency side, while for the space-only \ac{HPOD} occupies the negative-frequency side. The spatial spectral content, instead, provides only negative wavenumbers for the conventional \ac{HPOD} , while for the space-only \ac{HPOD} it provides only  positive wavenumbers. This means that the two implementations of the \ac{HPOD} deliver temporal and spatial modes, which are approximately the same apart from being one the complex conjugate of the other.

This behaviour can be explained by both the properties of the Hilbert transform (used to compute the analytic representation of the field) and the relation between the temporal frequency and spatial wavenumber provided by the phase velocity. The Hilbert transform operation used to complexify the flow field forces the negative side of the spectrum to zero in the direction in which the Hilbert transform is performed, i.e. for the conventional \ac{HPOD} performed in time the negative temporal frequency will be zero while for the space-only \ac{HPOD} performed in space the negative spatial wavelengths will be zero. Since temporal frequency and spatial wavenumber for a advective wave are linked through the phase velocity as $c=\omega/\lambda$, for advective waves moving towards positive $x$ direction, i.e. characterized by negative phase velocities, the temporal frequency of the space-only \ac{HPOD} has to be negative to compensate for the positive spatial wavenumber. Similarly, the conventional \ac{HPOD} needs to have negative wavenumbers, which manifest as a phase opposition in the spatial patterns with respect to the space-only \ac{HPOD}, characterized instead by positive wavenumbers. Notice that, both spatial and temporal modes delivered by the conventional and space-only HPOD are one the complex conjugate of the other. That is, when the spatiotemporal structure is assembled from these modes it behaves in the same physical way, making the 2 approaches equivalent for purely-advecting features.

\begin{figure}
\centering
\includegraphics[width=0.5\columnwidth]{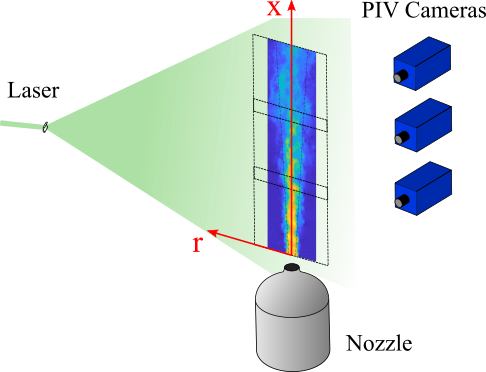}
\caption{A sketch of the turbulent-jet experimental setup.}
\label{img:setup}
\end{figure}

\subsection{Turbulent jet experiment}
As it has shown so far, both the conventional and space-only \ac{HPOD} approaches are capable of delivering modes, which can represent advecting features in the form of wavepackets. In most experimental settings, however, obtaining flow field data with enough temporal resolution to identify the underlying flow dynamics can be challenging. This lack of data rules out the application of the conventional \ac{HPOD} (as well as of other time-based data-driven decomposition techniques). Velocity field measurement techniques, however, provide instead enough spatial resolution to solve the wave displacement in the advection direction: this opens the way to the use of space-only \ac{HPOD} to extract advecting features from the data, even in lack of the dynamics information. To prove the suitability of this approach, the method is tested on snapshot \ac{PIV} data from a turbulent subsonic jet experiment described by \cite{raiola2019dynamic}. The jet issues from an axial-symmetric nozzle with exit diameter $D=20$\,mm with a jet exit velocity $V_J=25$\,m/s, resulting in a Reynolds number of $Re=V_J D/\nu=33\,000$. Planar \ac{PIV} is used to measure flow fields in the symmetry plane of the jet. 
Proper jet seeding is obtained by feeding into the stagnation chamber air premixed with DEHS droplets (with approximately 1\textmu m diameter) produced with a Laskin nozzle. The jet is confined in a closed structure with approximate dimensions $50D\times50D\times50D$ in order to have similar particle concentration inside and outside the jet. A double-pulsed Quantel Evergreen Nd:Yag Laser (200\,mJ/pulse at 15\,Hz) is used to illuminate a plane passing through the symmetry axis of the jet.
A set of 3  sCMOS Andor Zyla 5.5Mpixels cameras aligned along the jet axis direction is employed to image the flow field, spanning a field of view of  $6D \times 20D$. A sketch of the experimental setup is provided in Fig. \ref{img:setup}.  The particle images have been cross-correlated using a multi-pass image deformation algorithm. The final interrogation region size is $40\times40$ pixels with 75\% overlap, corresponding to a vector spacing of $0.029D$ in the velocity fields measured.
The final ensemble consists of 2600 velocity fields sampled at 10\,Hz repetition rate. In the present analysis, only half of the entire domain with respect to the jet centreline has been considered. This choice helps in preserving the circular symmetry of the problem and to avoid polluting the modes with information from the homogeneous azimuthal direction.

 \begin{figure*}
\centering
\includegraphics[width=\textwidth]{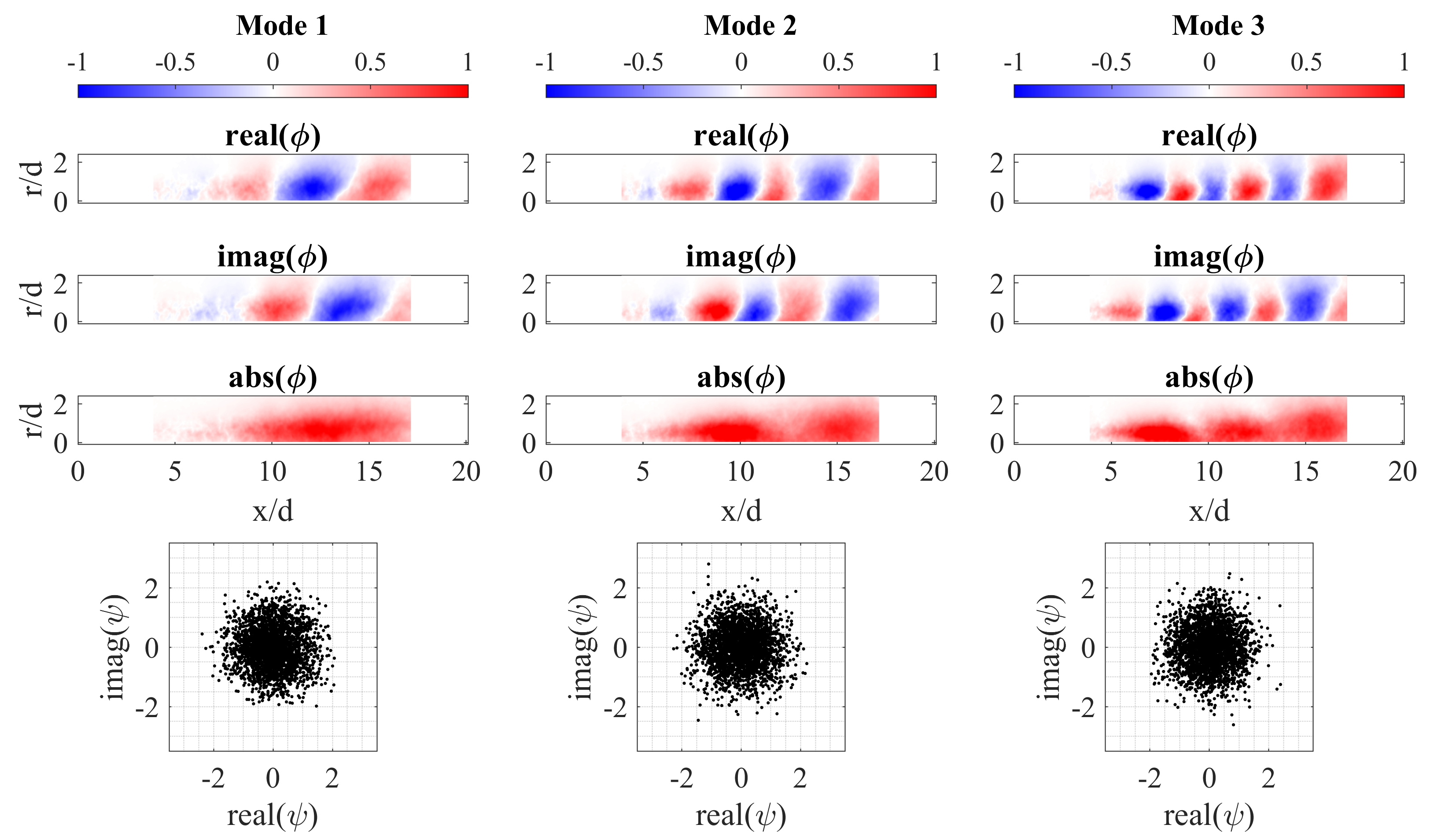}
\caption{Radial component of the complex-value \ac{HPOD} modes: (left) 1\textsuperscript{st} mode, (centre) 2\textsuperscript{nd} mode, (right) 3\textsuperscript{rd} mode. From top to bottom: real part of the spatial mode; imaginary part of the spatial mode; absolute value of the spatial mode; phase plot of the real versus imaginary part of the temporal mode. }
\label{fig:xHPOD_ExpJet_Modes}
\end{figure*}

The first 3 complex-valued structures obtained from the decomposition are depicted in Fig. \ref{fig:xHPOD_ExpJet_Modes}. These modes correspond to 8.8\%, 6.7\% and 5.4\% of the turbulent kinetic energy contained in the dataset, respectively. The complex-valued spatial modes $\phi$ are depicted both in their real and imaginary values in the first 2 rows. For each mode real and imaginary part have the same wavenumber content in space but shifted $\pi/2$ apart, thus indicating that they model a travelling wavepacket similarly to what has been observed for the LES case. Also, as the mode order is increased, their wavenumber increases. The absolute value of the mode, represented on the 3\textsuperscript{rd} row, reveals that each travelling wave undergoes a spatial amplification and decay, providing a spatial modulation of the wave in the streamwise direction which recall similar modulation patterns observed in the \ac{LES} case.

\begin{figure*}
\centering
\includegraphics[width=\textwidth]{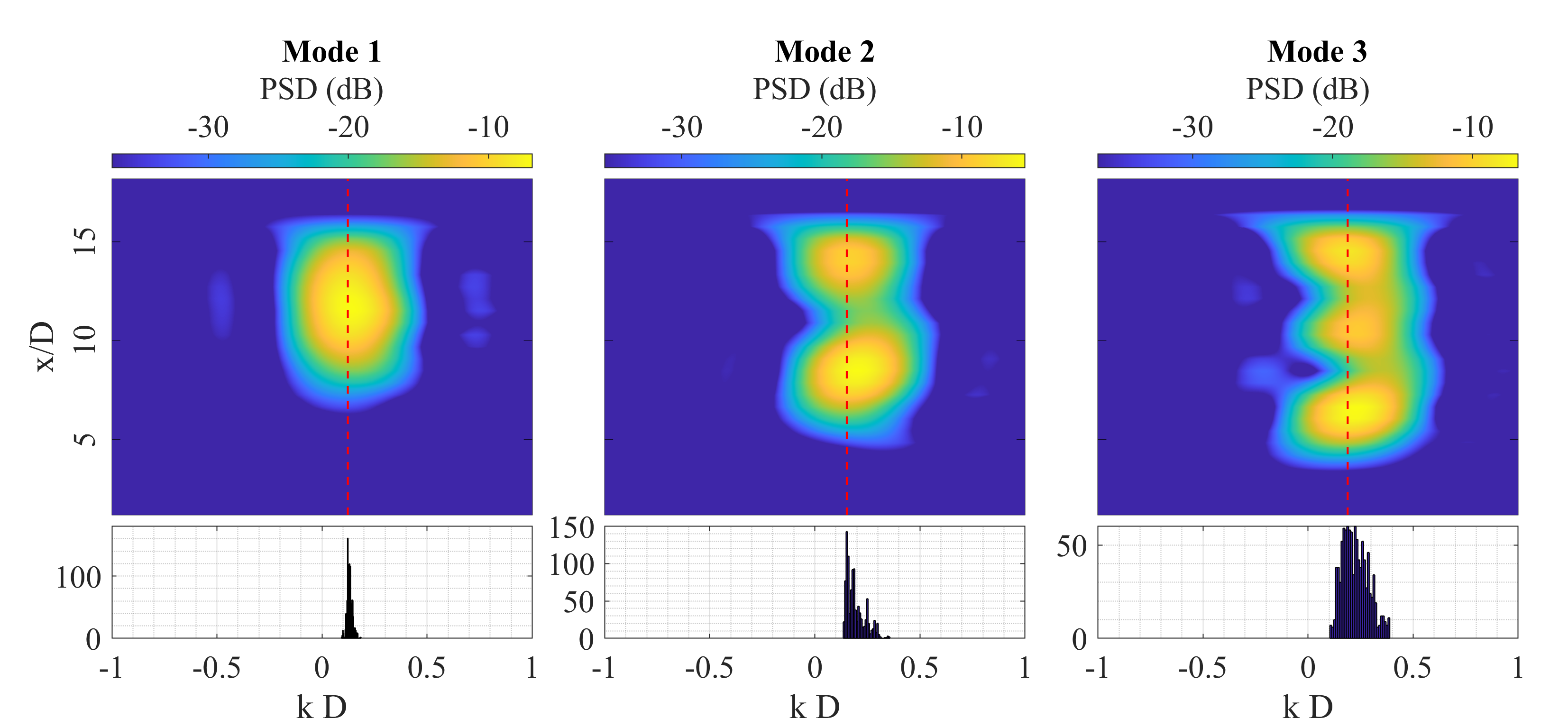}
\caption{Spectrogram along $x$ of the complex-valued temporal modes of the space-only \ac{HPOD}  (top) and histogram of the peak spatial frequency through the advective direction (bottom). From left to right: (left) 1\textsuperscript{st} mode, (centre) 2\textsuperscript{nd} mode, (right) 3\textsuperscript{rd} mode.}
\label{fig:xHPOD_ExpJet_SW}
\end{figure*}

The spatial spectral content of each spatial mode can be better appreciated in the space/wavenumber spectrograms in Fig. \ref{fig:xHPOD_ExpJet_SW}, obtained following the same procedure as for Fig. \ref{fig:xHPOD_Jet_LES_SW}. The analysis shows that -- as expected for the space-only HPOD -- the spatial spectra only include the positive-wavenumber side. The spectrograms for all modes are characterized by a strong modulation of the wave in amplitude and wavenumber along $x$. The spectral content is, however, quite band-limited, as already observed for the LES case: wavenumber modulation happens around a most probable value, as revealed by the peak-wavenumber histogram. Similarly to the LES case, this peak wavenumber value tends to increase as the mode number increases, capturing modes with decreasing energy content.

Finally, cyclograms of the real versus imaginary values of each temporal mode are depicted on the 4\textsuperscript{th} row of Fig. \ref{fig:xHPOD_ExpJet_Modes}. No clear unitary circle or other Lissajous figure can be spotted, suggesting that, similarly to what has been observed in the \ac{LES} case, the phenomenon is chaotic and dominated by modulation in time of the modes, which, for heavily undersampled data, would produce randomly filled circles in the cyclogram. 

\begin{figure}[tb]
\centering
\includegraphics[width=\textwidth]{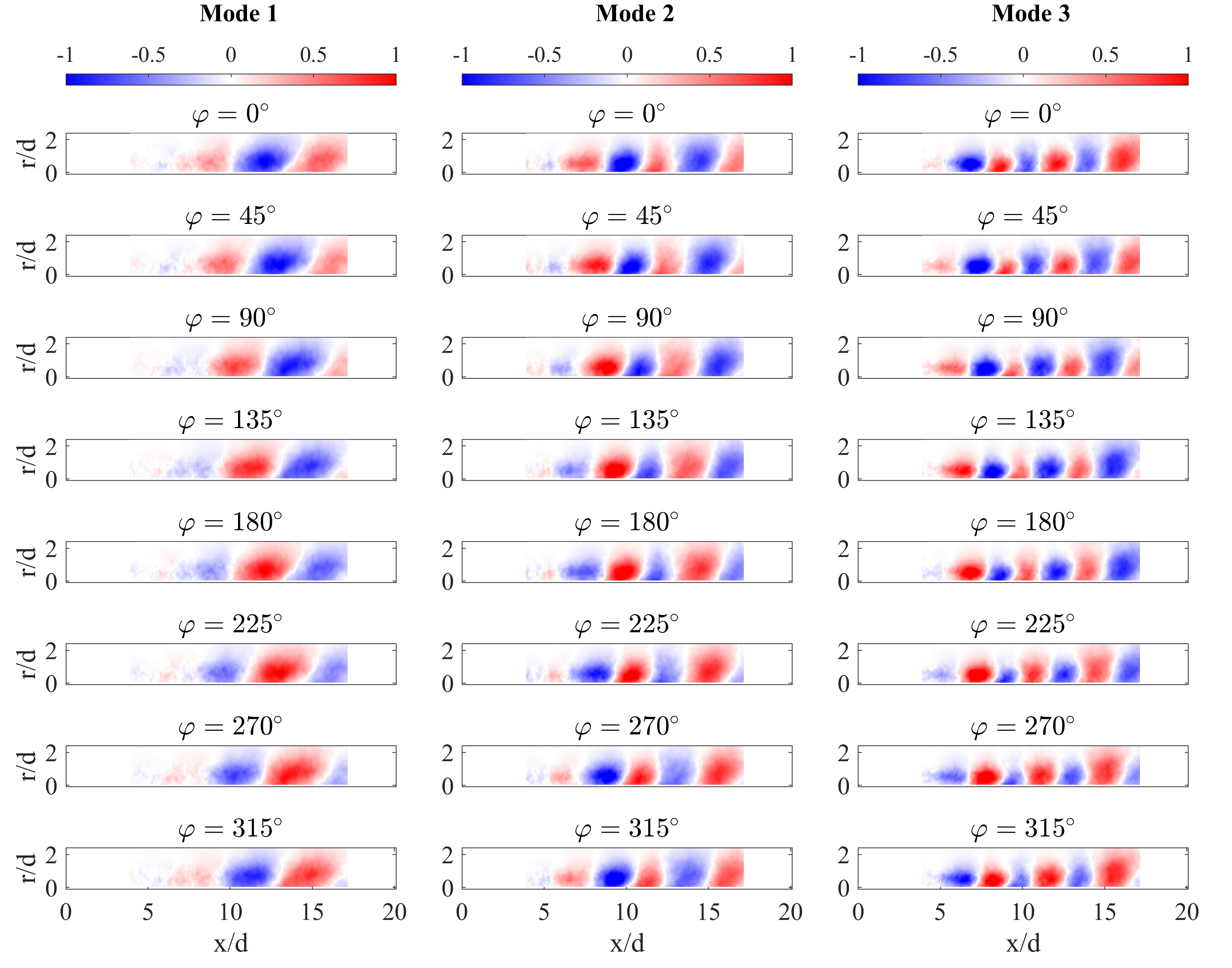}
\caption{Radial component of the oscillator model of the space-only \ac{HPOD} modes: (left) 1\textsuperscript{st} mode, (centre) 2\textsuperscript{nd} mode, (right) 3\textsuperscript{rd} mode. From top to bottom: different phases according to the oscillator model. }
\label{fig:xHPOD_ExpJet_Oscil}
\end{figure}

Despite the heavily undersampling of the PIV does not allow to extract the underlying dynamics of the flow, the space-only HPOD is capable of delivering spatial modes, which are compatible with advecting features. This can be better appreciated if an oscillatory model of these modes is produced by multiplying these complex-valued spatial modes by the complex exponential $\exp(i\varphi)$, where $\varphi$ is the temporal phase of the wave motion. Frames of the oscillatory model for the 1\textsuperscript{st}, 2\textsuperscript{nd} and 3\textsuperscript{rd} modes are provided in the 1\textsuperscript{st}, 2\textsuperscript{nd} and 3\textsuperscript{rd} columns of Fig. \ref{fig:xHPOD_ExpJet_Oscil}, respectively. The oscillatory model clearly shows the advecting behaviour underlying the spatial modes delivered by the space-only \ac{HPOD}, even if the actual behaviour in time -- likely characterized by modulation in frequency and amplitude -- cannot be extracted.

\section{Conclusions} \label{sec:concl}
This paper explores a complex-valued extension of the POD dubbed Hilbert POD (HPOD), and in particular introduces a novel space-only variant capable of retrieving spatiotemporally coherent wavepacket structures also in the absence of temporally resolved data. Even in its conventional version, this decomposition has found very little application in the field of fluid mechanics, so far. The present work aims at stressing the potential that this technique has, in particular regarding its capability to identify wavepacket structures in flows that are dominated by advective instabilities.

The mathematical framework supporting the HPOD is extensively discussed, in particular introducing the concept of analytic signal computed through the Hilbert transform. The analytic signal is shown to be capable of retrieving the complex-valued mathematical representation of a travelling wavepacket, expressed as a complex  exponential, from its real-valued part, which is the only one available from either measurements or simulations. Using this concept, the Hilbert POD is introduced. This decomposition, as the name suggests, is based on the complexification of the dataset through the analytic signal concept, prior to performing a standard POD. The conventional HPOD performs the Hilbert transform in the temporal direction. In this paper it is explored for the first time the possibility to compute the HPOD from the Hilbert transform performed in the spatial advection direction. This variant has been referred to as space-only, since it only requires spatial resolution in the data, oppositely to conventional HPOD where temporal resolution is required.
Using the mathematical framework introduced, it is demonstrated mathematically that the decomposition of the complexified dataset provides complex-valued temporal and spatial modes, which inherit the property of being analytic signals in the temporal or spatial direction, depending on which direction the Hilbert transform is performed into. These modes are therefore describing a complex exponential with local/instantaneous frequency and amplitude, thus representing the temporal or spatial function of a complex-valued wavepacket. It is worth remarking that, differently from other complex-valued decomposition such as the spectral POD, the spectral content of the modes has a local/instantaneous character and is not spectrally pure: this property allows the HPOD to deal with modulation and intermittency phenomena in the data at the cost of a less trivial description of the dataset.
Additionally, thanks to the similarity between advection time and space for a travelling wavepacket, it is demonstrated that computing the analytic signal in the advective direction or in time allows to extract equivalent spatiotemporally coherent structures: when a link between the temporal frequency and spatial wavenumber exists in terms of a phase velocity, the conventional HPOD will provide analytic spatial modes and the space-only HPOD would provide analytic temporal modes. This concept ensures that the retrieved structures are spatiotemporally coherent and opens the way to identify spatiotemporally coherent wave patterns also from non-time-resolved data using the space-only HPOD. Moreover, this concept can be used to discriminate between advective structures and other time-varying phenomena by using the most convenient direction to compute the analytic signal for the given flow scenario. Particularly, applying the space-only HPOD in the direction of propagation of a wave would allow to target more specifically said wave.

Both the conventional HPOD (in time) and the space-only HPOD (in the advective direction) are tested on 3 example problems in order to exemplify the properties of the decompositions and to test their performance.  

The first example problem is the velocity field in the wake of a cylinder at a Reynolds number of 100 obtained through DNS. This dataset is characterized by a vortex shedding with a pure shedding frequency and a compact reduced order model, as well as by the availability of temporal resolution. For this dataset it is shown that both implementations of the HPOD provide complex-valued modes, which represent the vortex shedding as wavepackets with constant frequency in time. The results are compared to standard real-valued POD, showing that for this very simple dataset the real and imaginary part of the HPOD modes correspond to subsequent pairs of real-valued POD modes. Additionally, this dataset is exploited to exemplify two other concepts: 1) signal conditioning, typically in the form of signal ends trimming, is required to remove edge effects introduced by the Hilbert transform operation and avoid corruption of the HPOD; 2) removing the temporal coherence from the dataset by shuffling the original time series does not allow the computation of conventional HPOD but do not alter the results of the space-only variant. This shows that both conventional and space-only \ac{HPOD} -- differently from space-only \ac{POD} -- can automatically deliver spatiotemporally coherent structures, without needing to pair together modes. The space-only \ac{HPOD} variant, in particular, can be adopted also in datasets with reduced temporal resolution. One obvious application of this space-only variant, therefore, is on snapshot PIV dataset, which typically offer great spatial resolution but do not allow to gather temporal information. Note however, that even though the conventional \ac{HPOD} fails to recognize correctly the mode pairing, when applied in a low temporal resolution setting, it will safely revert to a solution equivalent to the standard \ac{POD} one.

The second example problem is the time-resolved velocity field of a turbulent jet at a Reynolds number of $10^6$ obtained through LES. Differently from the first example, this dataset is dominated by wavepackets happening at multiple scales and characterized by a complex dynamics involving modulation phenomena and intermittency. Both variants of HPOD applied to this dataset retrieves practically identical complex-valued modes in time and space, which represent wavepackets. The spatial modes are characterized by a wave-like pattern in the advective direction, and amplitude and frequency modulation in the same direction. The temporal patterns are similarly characterized by modulation of amplitude and frequency as well, and clearly show the presence of intermittency of the retrieved wavepackets. Both spatial and temporal modes are analytic, i.e. are characterized by a spectral content limited only to the positive or negative side of the frequency spectrum. The sign of the spatial wavenumber and temporal frequency of the modes is compatible with a phase velocity of the wavepacket directed in the advection direction. Despite the spatial and temporal modes are not spectrally pure, the space-wavenumber and time-frequency analysis reveals that the spectral content is narrow-banded, further reinforcing the idea that the \ac{HPOD} targets advective structures. The capability of the \ac{HPOD} to absorb modulation effects, sets the present techniques apart from other techniques employed to identify advecting structures: differently from \ac{SPOD}, where the temporal frequency of each spatiotemporal structure is fixed on an infinite temporal support, the \ac{HPOD} provides spatiotemporal structures with instantaneous frequency. As such, the difference between the two decompositions is akin to the difference between Fourier analysis and time-frequency analysis, making them suited for different scopes. While the \ac{SPOD} can offer a clear and simple overview of flow behaviour in the space of frequencies, the \ac{HPOD} trades this simpler description of the flow to offer the capability of capturing spatiotemporally coherent wavepackets, whose frequency/wavenumber and amplitude are both local and instantaneous. This capability, for example, could be exploited to gather further insights in phenomena such as modulation and intermittency of turbulent flow features.

The third example problem is the velocity field of a turbulent jet at Reynolds number $33\,000$ measured by means of snapshot 2D PIV. This dataset represents a similar flow problem than the previous dataset, characterized by multiple scales and complex dynamics. The main difference is represented by the source of the dataset, which removes the availability of temporal resolution and introduces measurement uncertainty to the data. In this setting, the space-only HPOD is shown to retrieve spatial modes very similar to the ones identified in the LES dataset, confirming the capability of this variant to provide a physically meaningful decomposition even without accessing to temporal resolution. 
Indeed, the space-only HPOD can only provide a clear picture of the spatial modes, while the temporal ones do not allow to extract further information due to lack of temporal resolution. This spatial information, however, can still be especially valuable when paired with techniques for extracting the flow dynamics. For simpler flows, characterized by a spectrally-pure temporal frequency, oscillators models \citep{nair2018networked} can be built directly from the \ac{HPOD} modes. For more complex flows, techniques developed for the space-only \ac{POD}, such as Galerkin-POD projections models \citep{rowley2004model,jaunet2016pod} or techniques based on the correlation with high-repetition-rate sensors (see POD-based stochastic estimators such as the Extended POD, \citealt{boree2003extended,tinney2008low,hosseini2015sensor,discetti2018estimation}), can be extended to the \ac{HPOD} to provide more robust results in identifying a flow dynamic model.

\bibliographystyle{jfm}
\bibliography{bib}

\end{document}